\documentclass[sigconf]{acmart}

\newcommand{\algoName}[0]{\textsc{ArctyrEX}\xspace}

\setcopyright{none} 
\acmConference[Anonymous Submission to ACM CCS 2023]{ACM Conference on Computer and Communications Security}{Due 15 May 2023}{London, TBD}
\acmYear{2023}

\usepackage{filecontents}
\usepackage{amsmath}
\usepackage{subcaption}
\usepackage{stfloats}
\usepackage{verbatim} 
\usepackage{booktabs}
\usepackage{tkz-euclide}
\usepackage{textcomp}
\usepackage{xcolor}
\usepackage[utf8]{inputenc}
\usepackage{hyperref}
\usepackage{cleveref}
\usepackage{booktabs}
\usepackage{listings}
\usepackage{xcolor}
\usepackage{circuitikz}
\usepackage[linesnumbered,ruled,vlined, noend]{algorithm2e}
\ctikzset{logic ports=ieee}
\usetikzlibrary{positioning, fit, calc}
\usepackage{comment}
\usepackage{xspace}
\usepackage{soul}
\setstcolor{red}

\usepackage[T1]{fontenc}
\usepackage[scaled=0.8]{beramono}
\definecolor{light-gray}{gray}{0.45}
\definecolor{dark-green}{rgb}{0.0,0.34,0.25}

\usepackage{flushend}

\begin{document}
\title{\algoName : Accelerated Encrypted Execution \\of
General-Purpose Applications}

\author{Charles Gouert}
\affiliation{\institution{University of Delaware}}

\author{Vinu Joseph}
\affiliation{\institution{NVIDIA}}

\author{Steven Dalton}
\affiliation{\institution{NVIDIA}}
  
\author{Cedric Augonnet}
\affiliation{\institution{NVIDIA}}

\author{Michael Garland}
\affiliation{\institution{NVIDIA}}

\author{Nektarios Georgios Tsoutsos}
\affiliation{\institution{University of Delaware}}

\begin{abstract}
Fully Homomorphic Encryption (FHE) is a cryptographic method that guarantees the privacy and security
of user data during computation. FHE algorithms can perform unlimited arithmetic computations directly on encrypted data without decrypting it. Thus, even when processed by untrusted systems, confidential data is never exposed. In this work, we develop new techniques for accelerated encrypted execution and demonstrate the significant performance advantages of our approach.
Our current focus is the Fully Homomorphic Encryption over the Torus (CGGI) scheme, which is a
current state-of-the-art method for evaluating arbitrary functions in the encrypted domain. CGGI
represents a computation as a graph of homomorphic logic gates and each individual bit of the plaintext is
transformed into a polynomial in the encrypted domain. Arithmetic on such data becomes very
expensive: operations on bits become operations on entire polynomials. Therefore, evaluating even relatively
simple nonlinear functions, such as a sigmoid, can take thousands of seconds on a single CPU thread.
Using our novel framework for end-to-end accelerated encrypted execution called \algoName, developers with no knowledge of complex FHE libraries can simply describe their computation as a C program that is evaluated over $40\times$ faster on an NVIDIA DGX A100 and $6\times$ faster with a single A100 relative to a 256-threaded CPU baseline.
\end{abstract}

\maketitle

\section{Introduction} \label{sec:intro}

%
Cloud computing allows users to forego the practice of maintaining costly data centers in house, and provides both computation and storage capabilities on-demand. 
However, all user data will necessarily reside on servers owned by the cloud service provider who could view the uploaded data. 
Additionally, attackers are increasingly targeting cloud servers because they contain sensitive data from multiple users \cite{singh2017cloud} \cite{ali2015security} \cite{xiao2016one}. 
FHE allows users to encrypt data locally, outsource the ciphertexts to the cloud for oblivious and meaningful computation, and receive the encrypted processed data for decryption. 
This can be used for a wide variety of applications, such as privacy-preserving machine learning as a service (MLaaS) \cite{folkerts2021redsec} \cite{chou2018faster} \cite{chen2018logistic} and facial recognition \cite{yang2022design} \cite{boddeti2018secure}. 

FHE was realized in 2009 with the advent of the bootstrapping procedure which allows unlimited computation on ciphertexts \cite{gentry2009fully}. 
However, early FHE was plagued by both high memory requirements and enormous computational overheads, which rendered it infeasible for adoption. 
Since its inception, great strides have been made to reduce these runtime costs: 
First, new homomorphic encryption schemes have been developed with more efficient bootstrapping constructions, such as DM \cite{ducas2015fhew} and CGGI \cite{chillotti2020tfhe}. 
Additionally, various algorithmic and software optimizations, such as HE-friendly number theoretic transforms (NTT) \cite{dai2015cuhe}, have yielded significant speedups in encrypted computation for certain core operations. 
Additionally, utilization of the residue number system (RNS) has been employed to enhance parallelism and avoid large integer arithmetic \cite{halevi2019improved} \cite{cheon2019full} \cite{bajard2016full}.
Lastly, CPU-based acceleration techniques were also adopted, including AVX and FMA extensions \cite{boemer2021intel}.
However, the algorithmic level performance gains have recently stagnated and further speedups are coming only from hardware acceleration. 


The most prominent hardware platforms for encrypted computation with FHE are GPUs, which have been thoroughly demonstrated to be particularly suited for the types of arithmetic required by modern FHE constructions. 
Most encrypted operations expose ample parallelism and are computationally intensive \cite{kim2022secure}; therefore, FHE applications can leverage the high degrees of parallelism afforded by these devices. 
For instance, a $10\times10$ matrix multiplication in the encrypted domain using the CGGI cryptosystem in gate bootstrapping mode \cite{chillotti2020tfhe}, requires hundreds of millions of large polynomial arithmetic operations and NTTs.
%
Open-source nuFHE \cite{nufhe} and cuFHE \cite{cufhe} libraries expose an API akin to an assembly language, requiring programmers to compose their algorithms as Boolean circuits and their goal was to maximize the performance of individual homomorphic operations, as opposed to end-to-end encrypted applications themselves.
\newsavebox{\dotproduct}
\begin{lrbox}{\dotproduct}
\begin{lstlisting}[language=C,
    breaklines=true,         
    basicstyle=\ttfamily\footnotesize,
    numbers=left,
    numberstyle=\scriptsize\color{gray},
    keywordstyle=\color{blue},
    commentstyle=\itshape\color{gray}
]
int dot_product(int x[500], int y[500]) {
  int product = 0;
  for (int i = 0; i < 500; i++)
    product = product + x[i] * y[i];
  return product;
}
\end{lstlisting}
\end{lrbox}

\newsavebox{\fclayer}
\begin{lrbox}{\fclayer}
\begin{lstlisting}[language=C,
    breaklines=true,         
    basicstyle=\ttfamily\footnotesize,
    numbers=left,
    numberstyle=\scriptsize\color{gray},
    keywordstyle=\color{blue},
    commentstyle=\itshape\color{gray}
]
void fc_layer(short x[256], 
              short w[256], 
              short res[30]) {
  for (int i = 0; i < 30; i++) 
    for (int j = 0; j < 256; j++) 
        res[i] = res[i] + x[j] * w[j];
}
\end{lstlisting}
\end{lrbox}


\begin{figure*}[!t]
    \centering
    \begin{subfigure}[t]{0.5\textwidth}
        \centering
        \usebox{\dotproduct}
        \vspace{5pt}
        \caption{Dot Product Code}\vspace{0.1in}
    \end{subfigure}%
    ~
    \begin{subfigure}[t]{0.5\textwidth}
        \centering
        \usebox{\fclayer}
        \caption{Fully-Connected Layer Code}\vspace{0.1in}
    \end{subfigure}
    \begin{subfigure}[b]{0.5\linewidth}
        \centering
        \includegraphics[height=1.25in]{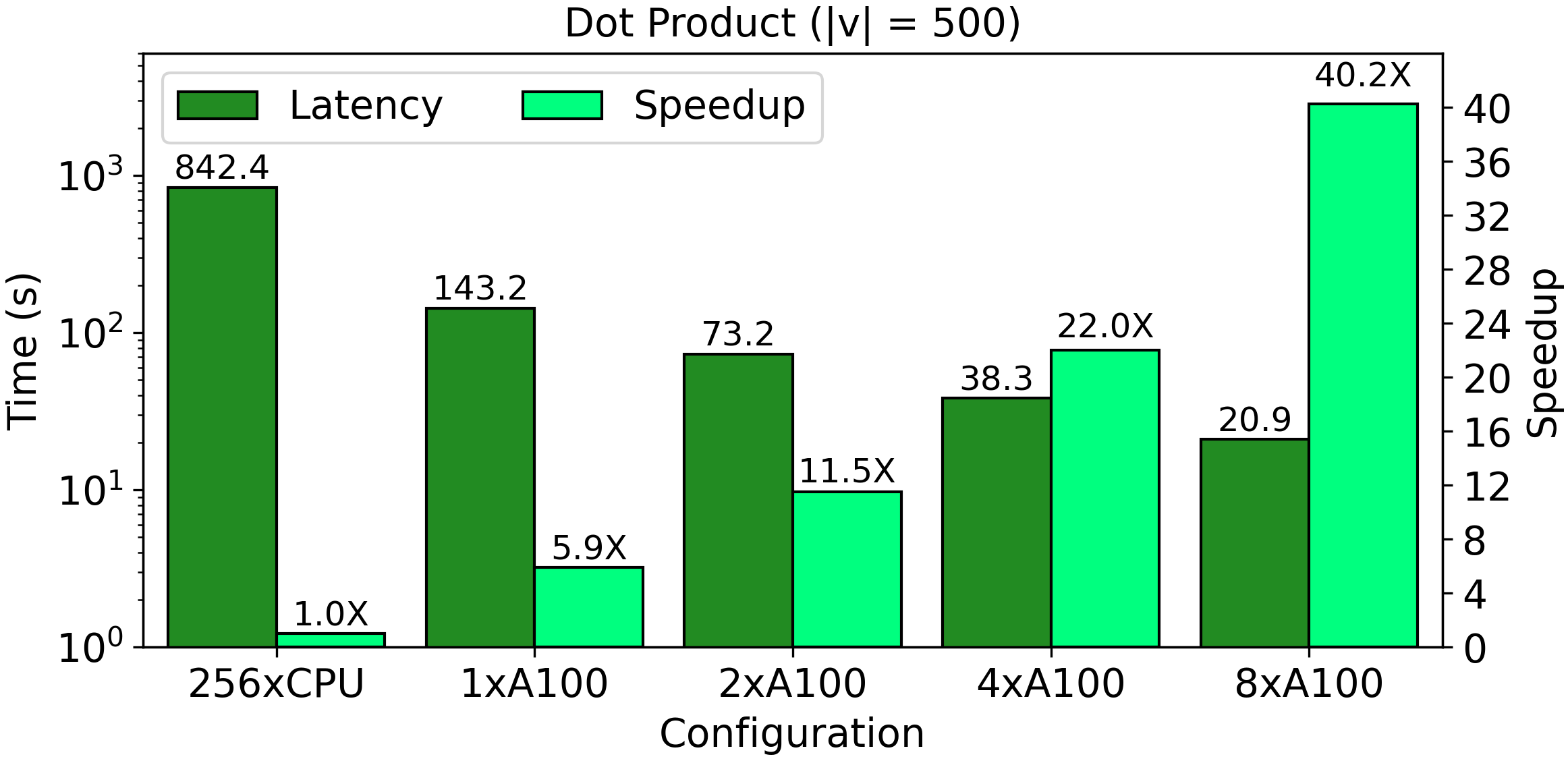}
        \label{f:dot}
        \caption{Dot Product Performance}
    \end{subfigure}%
    ~
    \begin{subfigure}[b]{0.5\linewidth}
        \centering
        \includegraphics[height=1.25in]{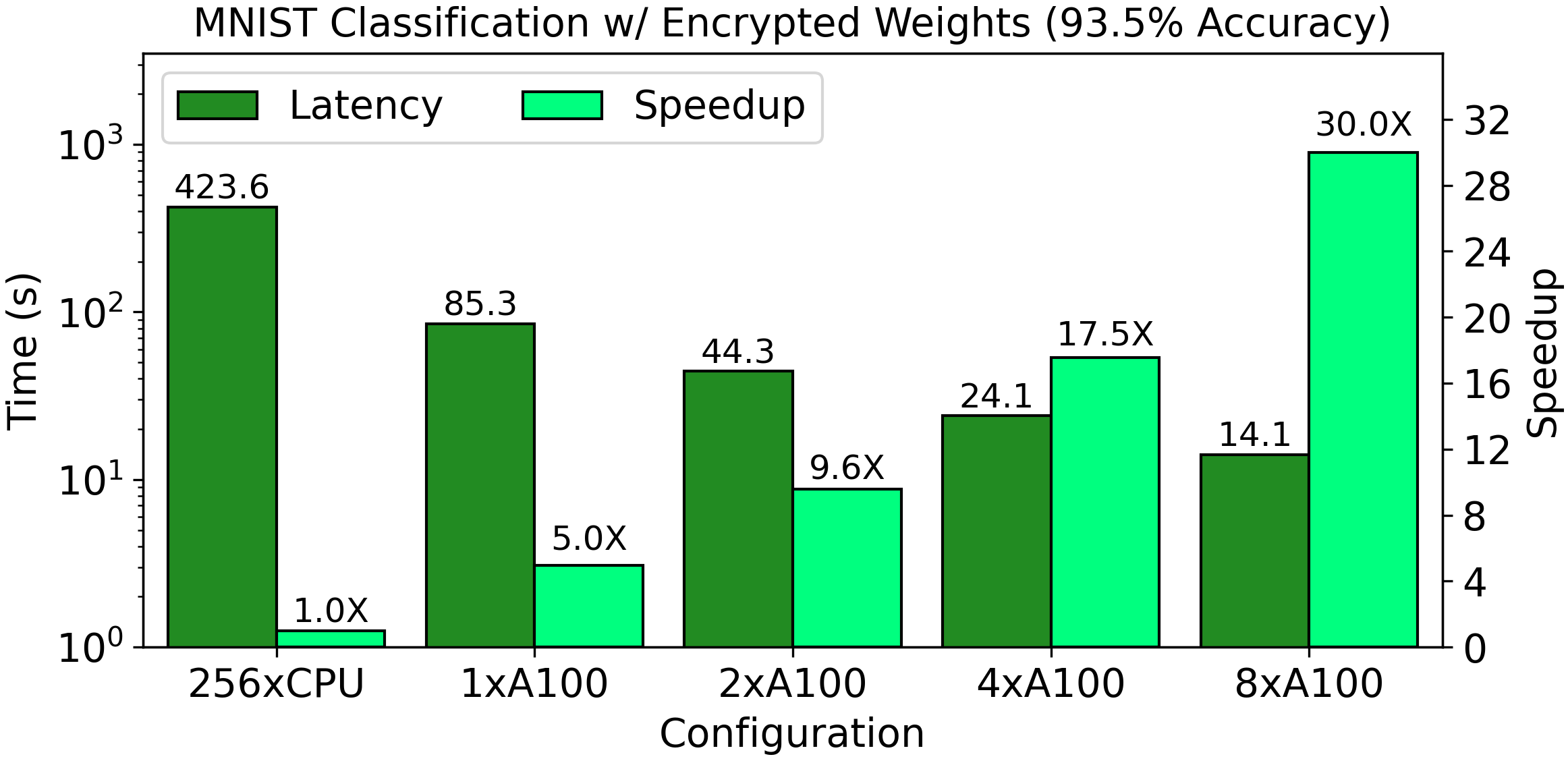}
        \label{f:interference}
        \caption{MNIST Classification Performance}
        \end{subfigure}
    \caption{\textbf{Practical Large-Scale Applications}:  Using our approach, a dot product subroutine runs $\approx 6\times$ faster on a single A100 GPU and over $40\times$ faster with an NVIDIA DGX A100 relative to a multi-threaded CPU execution with 256 threads, resulting in an end-to-end application level speed up of $30\times$ for MNIST classification.}    
    \label{f:exciting}\vspace{-0.1in}
\end{figure*}

In this work, we demonstrate that GPU-accelerated FHE can be used to greatly improve the efficiency of realistic and representative FHE applications, such as neural network inference and large linear algebra arithmetic. 
We also introduce automated scheduling techniques that allow for strong scalability while evaluating encrypted algorithms with multiple GPUs.
Notably, most cryptographic details and all hardware acceleration functionalities are handled automatically by \algoName to minimize the burden on programmers. 
Our key contributions can be summarized as follows:

\begin{itemize}
    \item A custom algorithm to translate high-level code to GPU-friendly FHE programs that reduces latency by up to 36\%, while also reducing circuit size by up to 40\% relative to a standard synthesis flow;
    
    \item A novel scheduling methodology that facilitates efficient computation across multiple GPUs, which enables encrypted programs to run up to $40\times$ faster on 8 GPUs;
 
    \item A new optimized backend for the CGGI cryptosystem that prioritizes fast evaluation of arbitrary algorithms and outperforms state-of-the-art implementations by more than an order of magnitude. This enables $32$-element vector addition of $32$-bit integers up to $4.1\times$ faster, and $16\times16$ matrix multiplication of $32$-bit elements up to $10.6\times$ faster on 8 GPUs.
    
\end{itemize}

Our proposed framework makes large-scale applications practical in FHE.
~Figure \ref{f:exciting}(a) showcases the high-level input code used to run both a large dot product of two vectors as well as a fully-connected layer in machine learning applications.
The user of our system simply needs to describe their computation as a C program; no knowledge of complex FHE libraries is required, except for the desired level of security. 
For C code outlining a dot product of two encrypted vectors of length 500, our framework automatically generates a highly efficient circuit consisting of 922308 gates with 128 levels resulting in approximately one billion combined NTT and inverse NTT invocations.

\section{Preliminaries} \label{sec:prelims}
This section discusses different variants of homomorphic encryption and provides the motivation for adopting fully homomorphic encryption for general-purpose computation. 
Additionally, it provides theoretical details regarding the CGGI cryptosystem employed in this work. 

\subsection{Homomorphic Encryption}
All encryption schemes that exhibit homomorphic properties enable meaningful computation directly on ciphertext data without revealing the underlying plaintext. 
The two variants of homomorphic encryption that support functionally complete sets of operations include leveled homomorphic encryption (LHE) and FHE. 
In both cases, ciphertexts are encoded as tuples of high-degree polynomials and adding or multiplying ciphertexts takes the form of polynomial addition or multiplication. 
These polynomials typically range from degree $2^{10}$ to $2^{17}$ and the coefficients are integers modulo $q$, which is a product of primes and typically hundreds of bits in length.
In the encrypted domain, addition increases the ciphertext noise slightly, while multiplication is significantly more noisy. 
An unfortunate consequence of this ciphertext noise (which is necessary for security) is that the noise magnitude increases during each homomorphic arithmetic operation, and eventually the noise will corrupt the underlying plaintext message and prevent successful decryption with high probability.
LHE can mitigate noise for a finite number of operations using a technique called \emph{modulus switching}, with larger encryption parameters allowing higher noise tolerance. 
However, larger parameters entail slower computation and higher memory consumption, which limits scalability for very deep circuits. 

FHE solves the scalability issues inherent to LHE and allows for unbounded, arbitrary computation on encrypted data. 
First realized by Gentry in 2009 \cite{gentry2009fully}, \emph{bootstrapping} is a noise mitigation technique that can be applied an infinite number of times, unlike modulus switching. 
In fact, any LHE scheme can be converted to an FHE scheme with the inclusion of bootstrapping. 
Nevertheless, the bootstrapping procedure itself is costly in terms of latency and remains a key bottleneck of all FHE constructions. 
Depending on the cryptosystem and chosen parameters, bootstrapping can take anywhere from several milliseconds \cite{chillotti2020tfhe} to minutes \cite{gentry2012better}. 
Therefore, the only way to achieve feasible FHE for general-purpose computation is to accelerate and optimize the bootstrapping mechanism.






\subsection{The CGGI Cryptosystem}

\tikzstyle{greentext} = [
            rectangle,
            minimum width=6.1cm,
            minimum height=0.8cm,
            text centered,
            fill=green!30]
\tikzstyle{yellowtext} = [
            rectangle,
            minimum width=6.1cm,
            minimum height=0.8cm,
            text centered,
            fill=yellow!30]
\tikzstyle{redtext} = [
            rectangle,
            minimum width=6.1cm,
            minimum height=0.8cm,
            text centered,
            fill=red!30]

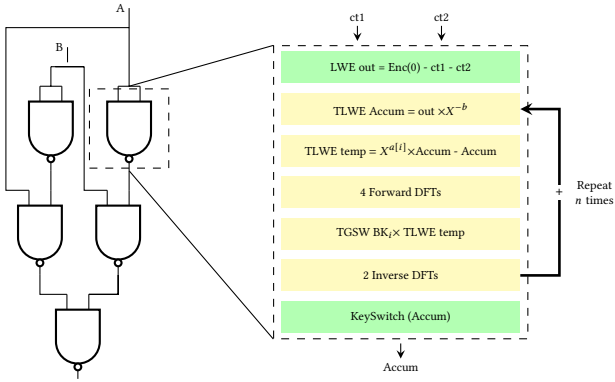
\begin{figure}
\centering
\begin{tikzpicture}[scale = 0.8]
    [
        dot/.style = {fill, shape=circle, minimum size=5pt, inner sep=0pt},
        rectext/.style= {rectangle, minimum width=6.1cm, minimum height=0.8cm, text centered}
    ]
\begin{scope}[rotate=-90,transform shape,scale=0.65]
\draw
(0,0) node[nand port](nand1){}
(nand1.in 1) -- (nand1.in 2) node[, midway](nand1_halfway){}
(nand1_halfway.center) -- ++(-1.5,0) node[](A){}
(A.center) -- ++(-0.5,0) node[left, rotate=90](A_label){A}

(nand1.out) to +(0.5,0) node[nand port, anchor=in 1](nand2){}
(nand2.out) -| +(0.5,-0.75) node[nand port, anchor=in 1](nand3){}
(nand2.out) -| (nand3.in 1)

(0,-2) node[nand port](nand4){}
(nand4.in 1) -- (nand4.in 2) node[midway](nand4_halfway){}
(nand4_halfway.center) -| ++(-0.5,0) node[](B_bottom){}
(B_bottom) -- ++(0,0.85) node[](B_top){}
(B_bottom.center) -- (B_top.center) node[midway](B_midway){}
(B_midway) -- ++(-0.5,0) node[left,rotate=90](B_label){B}

(nand4.out) to +(0.5,0) node[nand port, anchor=in 1](nand5){}
(nand5.out) -| (nand3.in 2)

(-1.5, -3.15) node[](lower){}
(A.center) |- (lower.center) -| (nand5.in 2)
(B_top.center) -| (nand2.in 2)
;
\end{scope}

\node [draw, dashed, rectangle, scale=5] at (nand1) (nand1_circle){};

\begin{scope}[transform shape, scale=0.65]
\node [greentext, below right=0.5cm and 4cm of A]       (node1) {LWE out = Enc(0) - ct1 - ct2};
\node [yellowtext, below=0.25cm of node1](node2) {TLWE Accum $=$ out $\times X^{-b}$};
\node [yellowtext, below=0.25cm of node2](node3) {TLWE temp $=X^{a[i]}{\times}$Accum - Accum};
\node [yellowtext, below=0.25cm of node3](node4) {4 Forward DFTs};
\node [yellowtext, below=0.25cm of node4](node5) {TGSW BK$_i \times$ TLWE temp};
\node [yellowtext, below=0.25cm of node5](node6) {2 Inverse DFTs};
\node [greentext, below=0.25cm of node6](node7) {KeySwitch (Accum)};

\node [draw, dashed, rectangle, minimum height=7.5cm, minimum width=6.5cm] at (node4) (list_rect){};

\draw [-stealth,line width=0.35mm](node6.east) -- ++(1,0) |- node[pos=0.25, fill=white](plus_sign){$+$}(node2.east);
\draw (nand1_circle.north) -- (list_rect.north west);
\draw (nand1_circle.south) -- (list_rect.south west);

\node [right=0.05cm of plus_sign, align=center, text width=1.05cm]{Repeat $n$ times};

\path (list_rect.north west) -- node[pos=0.33](input1){} node[pos=0.66](input2){}(list_rect.north east);
\draw [stealth-](input1) -- ++(0,0.5) node[above]{ct1};
\draw [stealth-](input2) -- ++(0,0.5) node[above]{ct2};

\path (list_rect.south west) -- node[pos=0.5](output3){} (list_rect.south east);
\draw [-stealth](output3) -- ++(0,-0.5) node[below]{Accum};
\end{scope}

\end{tikzpicture}\vspace{0.1in}

    \caption{\textbf{Encrypted Logic Gate Evaluation}: All standard two-input logic gates begin with a series of linear operations between the input LWE ciphertexts, followed by a bootstrapping procedure (executed by the looped instructions), and lastly a keyswitching operation. The steps in yellow represent operations associated with bootstrapping.}\label{f:tikz_logic_gate}
    \vspace{-0.2in}
\end{figure}

Both the DM \cite{ducas2015fhew} and CGGI cryptosystems \cite{chillotti2020tfhe} possess bootstrapping routines that can be evaluated in up to tens of milliseconds on a CPU using modern open-source implementations such as Concrete~\cite{chillotti2020concrete} and OpenFHE~\cite{al2022openfhe}, which is faster than other FHE cryptosystems.
Also, while other schemes encrypt vectors of integers and floating point numbers, CGGI and DM are typically used to encrypt individual bits into a single ciphertext. 
Due to this encoding, the core encrypted operations take the form of Boolean gates, which are more flexible in terms of general computation than arithmetic operations over integers (e.g., encrypted comparisons are easily implemented using encrypted bits). 

 As discussed, to support unlimited computation depths, the FHE scheme must periodically invoke a bootstrapping operation to reset the amount of noise in the ciphertext. In the case of CGGI, which evaluates Boolean gates, bootstrapping must be performed every gate. As a result, evaluating a single homomorphic gate requires on the order of 2,000 polynomial multiplications \cite{chillotti2020tfhe}, which are typically accomplished using the Discrete Fourier Transform (DFT). While this is an efficient algorithm for a single polynomial multiplication, even a small application could require billions of DFTs. For example, the computation graph for a single inner product of two vectors comprising 16 encrypted 16-bit numbers contains nearly 25,000 encrypted logic gates. Evaluating this circuit results in over 75 million invocations of the DFT.
DM \cite{ducas2015fhew} was the first cryptosystem to introduce a \textit{functional bootstrap} that refreshes noise while simultaneously evaluating a non-linear operation on the encrypted bits. 
In fact, this bootstrap is a necessary component of the computation for logic gates such as \texttt{NAND}.

CGGI improves upon this construction and generalizes it for all logic gates, including an encrypted \texttt{MUX} that is capable of obliviously choosing between two encrypted bits dependent on the underlying value of an encrypted selector bit. 
For all gates except the inverter, which is noiseless and linear, the bootstrapping operation comprises the majority of the gate's latency \cite{jiang2022matcha}. 
In turn, the core bottleneck of bootstrapping is the numerous polynomial multiplications between encrypted secret key components and input ciphertexts.
Most FHE libraries perform these high-degree polynomial multiplications as element-wise multiplications in the DFT domain, which is asymptotically faster than textbook polynomial multiplication \cite{chen2014high} \cite{cormen2022introduction}.
Both the number theoretic transform (NTT) and fast fourier transform (FFT) can facilitate the forward and inverse domain conversions for these purposes.
However, the NTT is typically preferred over the FFT as it operates directly over integers.
Moreover, FFT requires additional type conversions between integers and floating point numbers as FHE ciphertexts contain integer coefficients. As a result, FFT introduces small computation errors due to its reliance on floating point arithmetic.

The CGGI cryptosystem employs different types of ciphertexts, each with different characteristics. 
The first type, known as LWE ciphertexts, serve as the inputs and outputs of each homomorphic gate evaluation from a user perspective. 
LWE ciphertexts are the smallest type that CGGI uses; at 128 bits of security, they consist of a single 630-degree polynomial with 32-bit coefficients and an extra 32-bit scalar term. 
However, these ciphertexts can not be used for nonlinear encrypted operations and are incapable of being used to evaluate a standard encrypted logic gate function (with the trivial \texttt{NOT} gate being the sole exception). 
Instead, these ciphertexts need to be transformed to TLWE ciphertexts (i.e., Ring-LWE) that are larger in size.
Typically, TLWE ciphertexts are composed of a tuple of 1024-degree polynomials with 32-bit coefficients.
The third type is TGSW ciphertexts, which are the largest and can conceptually be viewed as an array of TLWE ciphertexts. 
The bootstrapping key, which is an encryption of the secret key, is composed of this type of ciphertexts. 
Importantly, TGSW ciphertexts can be multiplied directly with TLWE ciphertexts, which is a necessary step of all bootstrapped gate evaluations. 
Figure \ref{f:tikz_logic_gate} gives a high-level overview of the operations involved in a homomorphic \texttt{NAND} gate.
All bootstrapped gates are evaluated in a similar way and only differ in the preliminary linear operations (i.e., the top green box in the figure).

Overall, the CGGI cryptosystem is a good candidate for achieving accelerated general purpose computation on GPUs for a variety of reasons. 
First, the parameters used by CGGI are often significantly smaller than other FHE schemes, which yields smaller ciphertexts. 
Thus, multiple ciphertexts can be held in the GPU shared memory simultaneously, which is not always the case for schemes such as CKKS \cite{cheon2017homomorphic}, BFV \cite{fan2012somewhat}, and BGV \cite{brakerski2014leveled} that can utilize ciphertexts on the order of several megabytes \cite{varia2015hetest}.
Notably, certain classes of encrypted operations used for general purpose computation are well-suited for CGGI with binary ciphertexts, but are non-trivial using other cryptosystems that adopt multi-bit encodings. 
For instance, comparison operations, bitwise manipulations like shifting, and nonlinear functions such as the ReLU activation function in machine learning applications, can be computed directly without the need of costly polynomial approximations \cite{boura2019simulating} \cite{han2019logistic}. 
Lastly, the requirement of executing hundreds of DFT transforms per bootstrap is particularly well-suited to GPUs due to the parallel nature of FFT and NTT.
\textcolor{black}{CGGI can also support multi-bit encodings and employ a special programmable bootstrapping mechanism that evaluates univariate functions. However, only low precision is achievable with realistic parameters and therefore this approach is better suited for specific applications rather than for arbitrary computation.
We strictly consider CGGI in gate bootstrapping mode with binary ciphertexts in this work specifically for this reason but note that the methodology proposed is readily extensible to support this scenario.}

\section{\algoName : System Design} \label{sec:frontend}
\begin{figure*}[!ht]
    \centering
    \includegraphics[width=0.75\textwidth]{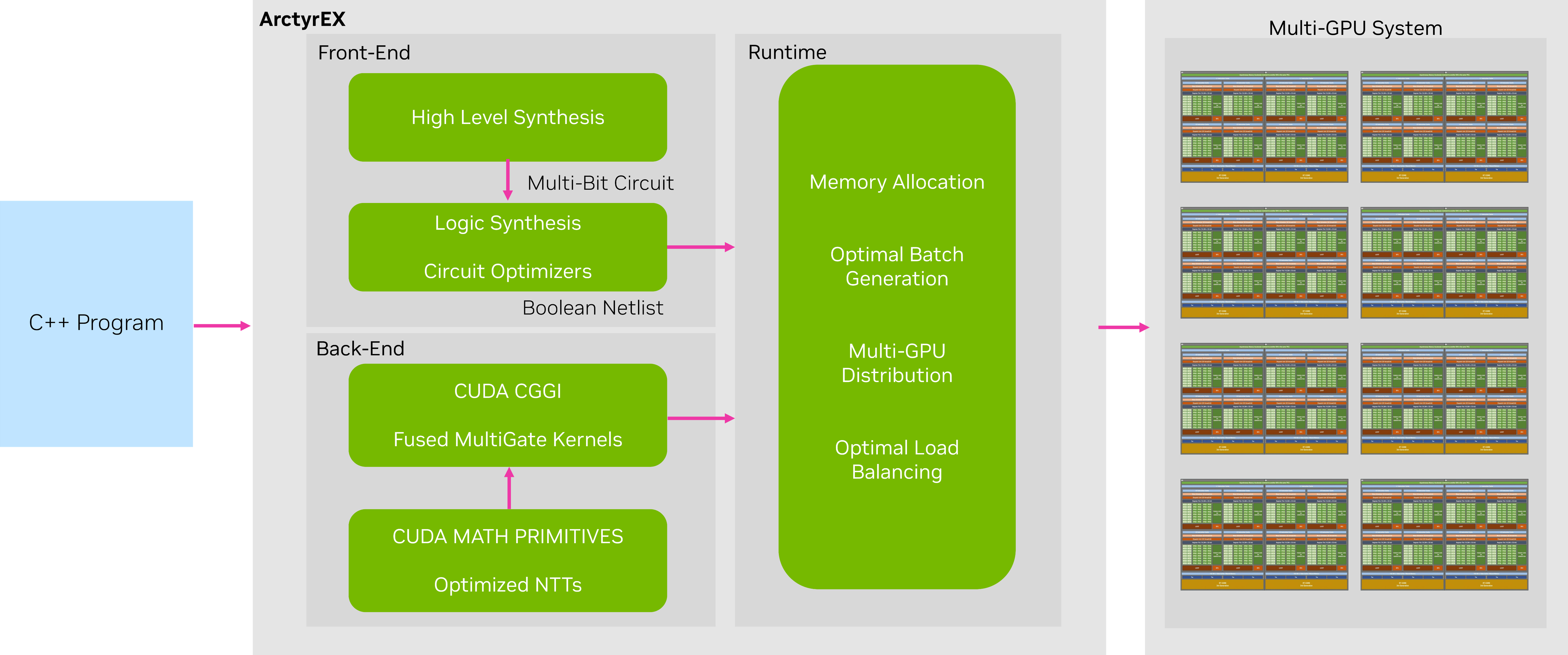}
    \caption{\algoName \textbf{System Overview:} Our proposed system is composed of three distinct layers that work together to realize an end-to-end framework for scalable encrypted computation. The frontend converts high-level programs to a logic circuit tuned for FHE. In turn, this logic circuit is parsed by the middle layer, which executes a coordination algorithm that partitions each level of the circuit into shares and assigns them to multiple GPUs. The back-end enables outsourcing computationally expensive FHE operations in each share to the GPUs.}\label{f:overview}
\end{figure*}

\algoName is an end-to-end framework that allows users to seamlessly convert high-level programs written in C to a sequence of GPU-friendly FHE Boolean operations leveraging the CGGI cryptosystem.
An overview of the system is depicted in Figure \ref{f:overview}, illustrating the capabilities of the frontend, runtime schedule coordination, and backend operations. 
Our proposed frontend tackles challenges associated with leveraging CGGI from a user perspective, such as adapting to the Boolean circuit model. 
%
In this section, we identify desirable circuit characteristics for efficient execution on GPUs and describe key aspects of the synthesis flow used to convert input programs to FHE code for outsourced computation.

\subsection{Optimal FHE Circuit Characteristics}

\newsavebox{\mmex}
\begin{lrbox}{\mmex}
\begin{lstlisting}[language=C,
    breaklines=true,         
    basicstyle=\ttfamily\footnotesize,
    numbers=left,
    numberstyle=\scriptsize\color{gray},
    keywordstyle=\color{blue},
    commentstyle=\itshape\color{gray}
]
void full_gemm(short x[100], short y[100], short res[100]) {
  for (int i = 0; i < 10; i++) {
    for (int j = 0; j < 10; j++) {
      res[10*i + j] = 0;
      for (int k = 0; k < 10; k++) {
        res[10*i + j] = res[10*i + j] + x[i*10 + k] * y[k*10 + j];
      } } } }
\end{lstlisting}
\end{lrbox}

\newsavebox{\lrex}
\begin{lrbox}{\lrex}
\begin{lstlisting}[language=C,
    breaklines=true,
    basicstyle=\ttfamily\footnotesize,
    numbers=left,
    numberstyle=\scriptsize\color{gray},
    keywordstyle=\color{blue},
    commentstyle=\itshape\color{gray}
]
int lr_inference(int data[4], int weights[4], int bias) {
  int product = 0;
  for (int i = 0; i < 4; i++)
    product = product + data[i] * weights[i];
  product = product + bias;
  // Sigmoid: 40320 + 20160*x - 1680*x^3 + 168*x^5 - 17*x^7
  int temp = 40320;
  int squared = product * product;
  temp += 20160 * product;
  temp -= squared * product * 1680;
  squared *= squared;
  temp += squared * product * 168;
  squared *= product * product;
  product = temp - squared * 17;
  // Client post-processes score by dividing by 80640
  return product;
}
\end{lstlisting}
\end{lrbox}

\begin{figure}[ht!]
\centering
\begin{subfigure}[b]{\textwidth}

\usebox{\mmex}
\end{subfigure}
\begin{subfigure}[b]{\textwidth}

\usebox{\lrex}
\end{subfigure}
\caption{HLS Kernels for General Matrix to Matrix Multiplication (GEMM) and Logistic Regression (LR) Inference.}\label{f:mmex}
\end{figure}

\begin{figure}[!ht]
    \centering
    \includegraphics[width=\columnwidth]{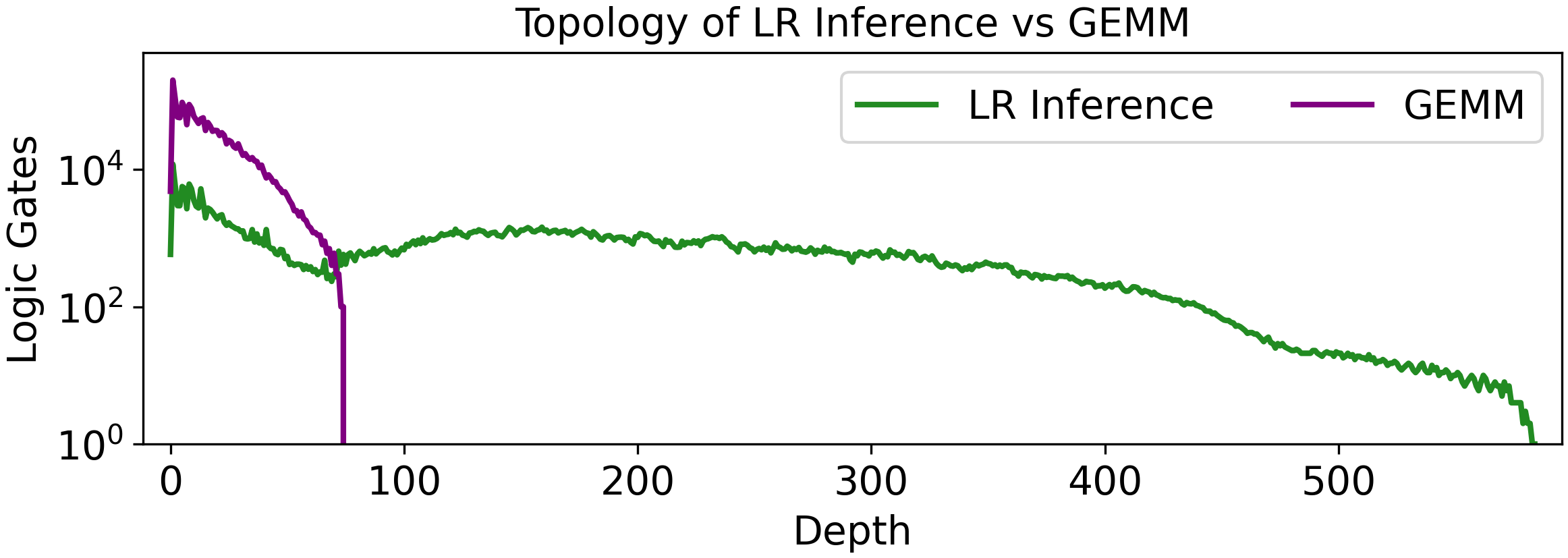}
    \caption{\textbf{Circuit-level parallelism}: Visualizing the circuit topology of GEMM versus LR inference shows that GEMM is ideal for parallel evaluation, while LR is less suitable.}\label{f:levels}
\end{figure}

One of the challenges for achieving efficient encrypted computation with the CGGI cryptosystem involves exploiting \textit{circuit-level parallelism} at the logic gate level. 
Essentially, any number of gates with resolved dependencies (e.g., all input wires have been loaded with encrypted ciphertexts) can be executed in parallel as they are entirely independent. 
For CPU-based systems with a limited number of cores, this parallelism is sufficient to effectively saturate the available CPUs without any significant optimizations at the logic synthesis or application level. 
However, high-performance computing systems that leverage hundreds of CPU cores or incorporate GPUs require much higher degrees of circuit-level parallelism to achieve high efficiency.
For these systems, the characteristics of the underlying Boolean circuit become much more important, therefore avoiding sub-optimal configurations is a critical concern. 
For example, using the kernels of Figure \ref{f:mmex}, we present the width of each circuit level for a $10\times10$  matrix multiplication as well as a logistic regression (LR) inference in Figure \ref{f:levels}.
The matrix multiplication benchmark represents ideal circuit characteristics for parallel execution as the majority of levels are very wide (the largest being nearly 200,000 gates in width), and the critical path is relatively short. 
On the other hand, LR inference has over 500 levels (resulting in a much longer critical path) and the width of each level is considerably shorter than those in the matrix multiplication circuit.
Another type of circuit configuration ill-suited for systems that can exploit high degrees of parallelism is circuits that adopt cascading. 
Cascaded circuits typically have a long critical path and each level of the circuit is narrow, limiting the number of gates that can be evaluated in parallel at any given time. 

Likewise, not all encrypted gates have the same execution time. For instance, \texttt{NOT} gates are significantly faster than other gates because they don't require any bootstrapping, while \texttt{MUX} gates are approximately twice as expensive as standard gates (like \texttt{AND} and \texttt{OR} gates). Efficient FHE circuit generation should take into account these differences in gate efficiency. 

\subsection{Synthesizing FHE-friendly Circuits}
The conversion process from a C program to an equivalent FHE algorithm can be completed in two distinct steps borrowed from modern hardware design paradigms: high-level synthesis (HLS) followed by logic or register transfer level (RTL) synthesis.  
While any HLS tool can be used for this purpose, we employ the Google XLS framework \cite{google_xls}, which is a fast and efficient open-source HLS engine that can be used to rapidly generate synthesizable Verilog code.
This Verilog code serves as an intermediate representation and describes the circuit functionality, which is then transformed by a logic synthesis tool to generate the actual Boolean netlist. 

We utilize the open-source Yosys Open Synthesis Suite to facilitate this process and perform crucial circuit-level optimizations \cite{yosys}. 
However, all existing logic synthesis tools, including Yosys, are tailored specifically for physical hardware development and optimize for several constraints that are not relevant to virtual FHE circuits (such as minimizing area or reducing clock cycle latency). 
The most relevant factors for optimal FHE circuit generation are minimizing the critical path delay, which is luckily a goal shared with actual hardware development, and prioritizing gates that run efficiently in the encrypted domain. 
Similarly to techniques introduced by the Google Transpiler \cite{gorantala2021general}, we can configure the logic synthesis tool to choose FHE-friendly gates by encoding the relative costs of each gate type as a function of area.
For instance, we assign the multiplexer gate to be twice as big as the standard two-input gates to reflect the fact that the latency of the \texttt{MUX} is twice as slow as a standard gate. 

Where prior work has adopted generic synthesis scripts for generating netlists for FHE evaluation \cite{gorantala2021general, gouert2020romeo}, our synthesis flow: (1) reduces the time required to generate the netlist relative to the Yosys generic synthesis script, and (2) results in more efficient circuits for FHE.
The core optimizations that we utilize with Yosys include functional and word-size reduction, removing redundant logic, and omitting unreachable branches in decision trees.
Compared to the baseline Google XLS logic optimizations, we observe a reduction of about 40\% in the overall size of the circuit for a dot product of two vectors with length 500.
However, we note that the Google XLS logic optimizer is more lightweight and can process the encrypted circuit about twice as fast.
We emphasize that this process is a one-time cost; after the circuit is processed, it can be executed using an arbitrary number of inputs.




\section{Novel Scheduling Algorithm for Scalable Evaluation} \label{sec:middleware}
The \algoName runtime library implements our proposed scheduler that allows homomorphic applications to utilize multiple computing resources with high scalability. 

\subsection{Strategies for Evaluating FHE Circuits}
After the Boolean netlist has been generated by the frontend compiler, and before encrypted computation can be carried out, we need to translate each gate to the encrypted domain. 
This process involves traversing the circuit, which is represented as a directed-acyclic graph (DAG), and mapping each node to the equivalent CGGI gate function. 
All wires become ciphertext data, the inputs are loaded with encryptions provided by the client, while the outputs are communicated back to the client for decryption after circuit evaluation. 

The intuitive approach for providing the computing party with an executable FHE circuit is to simply generate code that invokes the encrypted gate functions using the underlying backend directly one after the other. 
This approach works well for small programs where performance is not critical, but is ill-suited for non-trivial programs. 
For complex programs, the generated FHE code can grow to millions of lines in length, as each logic gate in the circuit would require 2-3 lines of code on average. 

Our key observation is that it is more efficient to avoid code generation entirely and incorporate a scheduler that traverses the DAG and distributes each gate to additional workers that exclusively run the corresponding FHE logic gate function. 
In this approach, gates that are ready to be evaluated can be distributed across a set of workers to exploit the circuit-level parallelism inherent in all applications. 
We remark that a similar methodology is employed by the Google FHE transpiler \cite{gorantala2021general} and is referred to as \emph{interpreter mode}. 
However, their strategy of distributing gates one at a time is not feasible when the workers constitute GPUs. 
Previous GPU-centric CGGI implementations as well as our proposed implementation (described in Section \ref{sec:backend}) can execute one homomorphic logic gate per streaming multiprocessor (SM) concurrently. 
In the case of an NVIDIA A100 GPU, $108$ homomorphic logic gates are the least number required to achieve $100\%$ device utilization at any given time. 
Thus, only one SM could be engaged if gates are assigned one at a time, resulting in extremely inefficient evaluation.
Further, interpreter mode creates new ciphertext objects and allocates more memory as needed throughout circuit evaluation. 
While this technique is suitable for CPU workers, it therefore creates a prohibitive bottleneck on GPUs as memory allocation and ciphertext transfers between the host and the device are costly. 

%

\subsection{\algoName Runtime Library}\label{ss:runtimelib}
We propose a novel methodology for efficient evaluation of encrypted circuits on both CPU and GPU devices. The host thread parses the intermediate representation (IR) generated by the frontend, and generates a set of nodes stored using XLS data structures~\cite{google_xls}. Each node contains an opcode, which defines the operation performed by the circuit gate, and its input operands, which are pointers to other XLS nodes.

The IR thus defines a sequence of gates which can be processed sequentially to generate a valid execution of the circuit. In order to introduce parallelism, we transform this ordered set of XLS nodes into a vector of circuit gates.
In addition to logic gates, we also create gates which compute encrypted constant values and augment the frontend IR with gates which copy an input ciphertext into another one. These copies are used to integrate the retrieval of encrypted results as part of the circuit itself instead of having to extract individual ciphertexts after waiting for circuit termination.

To derive a parallel execution of the circuit, we first dispatch all gates into multiple \emph{waves}. Each wave must be processed in-order, but all entries of a \emph{wave} can be processed concurrently. We now detail the topological sort algorithm that we use to build the list of \emph{waves}.

For each entry of the vector of gates, we compute its successors (gates depending on it), and count its predecessors (gates on which it depends).
To assign gates to different waves, we create a FIFO of ready gates, which are gates with no remaining dependencies. We start by adding all gates with no input dependencies into this FIFO. Until the FIFO is empty, we remove the first entry $n$ from the FIFO, and do the following :
\begin{itemize}
    \item Assign $n$ to the first wave if there are no input dependencies, or compute the max index of the wave of all predecessors, and add $n$ to the next wave. All predecessors have an index or $n$ would not be in the ready FIFO.
    \item We decrement the predecessor count of all successors of $n$. Any of these successors reaching a null predecessor count are put in the ready FIFO.
\end{itemize}

\begin{figure}[!ht]
\centering
\begin{circuitikz}[ieee ports, scale=0.2, transform shape]
\foreach \x in {3,...,0}{
    \ifnum\x=3\relax
        \draw node[nand port](P\x){} (P\x.in 1)
        to [short, -o] ++(-2, 0) coordinate(base_label) ++(-0.5, 0) node{$A\x$} (P\x.in 2)
        to [short, -o] ++(-2, 0) ++(-0.5, 0) node{$B\x$} ;
    \else
        \draw node[nand port, below=6cm of P\number\numexpr\x+1\relax](P\x){} (P\x.in 1)
        to [short, -o] ++(-2, 0) ++(-0.5, 0) node{$A\x$} (P\x.in 2)
        to [short, -o] ++(-2, 0) ++(-0.5, 0) node{$B\x$};
    \fi
    
    \draw node[and port, above right=0mm and 0.8cm of P\x, anchor=in 2](AT\x){};
    \draw node[and port, below=0.5cm of AT\x](AB\x){};
    \draw node[nor port, right=3 of P\x](BA\x){};
    \draw (AT\x.out) |- (BA\x.in 1) (AB\x.out) |- (BA\x.in 2);
}

\draw node[and port, above right=1.5 and 7 of P3, anchor=in 2](CT3){};

\foreach \x in {4,...,0}{
    \ifnum\x=4\relax
        \draw node[and port, below=3.5cm of CT3, number inputs=3](CP\x){};
    \else
        \ifnum\x=3\relax
            \draw node[and port, below=0.2 of CP\number\numexpr\x+1\relax, number inputs=4](CP\x){};
        \else
            \draw node[and port, below=0.2 of CP\number\numexpr\x+1\relax, number inputs=5](CP\x){};
        \fi
    \fi
}

\foreach \x in {9,...,5}{
    \ifnum\x=9\relax
        \draw node[and port, below=4.25cm of CP0, number inputs=5](CP\x){};
    \else
        \ifnum\x=5\relax
            \draw node[and port, below=0.2 of CP\number\numexpr\x+1\relax, number inputs=3](CP\x){};
        \else
            \ifnum\x=6\relax
                \draw node[and port, below=0.2 of CP\number\numexpr\x+1\relax, number inputs=4](CP\x){};
            \else
                \draw node[and port, below=0.2 of CP\number\numexpr\x+1\relax, number inputs=5](CP\x){};
            \fi
        \fi
    \fi
}

\draw node[and port, below=4.25 of CP5](CB3){};

\draw (CP0.in 5) -- coordinate[midway](cp0_cp9) (CP9.in 1);
\draw node[and port, right=3.5cm of cp0_cp9, number inputs=5](DB1){} (DB1.out) to [short, -o, l={$A=B$}] ++(0.1, 0);
\draw node[nor port, above=3.75cm of DB1, number inputs=6](DB0){} to (DB0.out) to [short, -o, l=$A>B$] ++(0.1, 0);
\draw node[nor port, below=3.75cm of DB1, number inputs=6](DB2){} to (DB2.out) to [short, -o, l=$A<B$] ++(0.1, 0);

\draw (CT3.out) -| (DB0.in 1)
      (CP4.out) -| ++(0.95, 0) |- (DB0.in 2)
      (CP3.out) -| ++(0.75, 0) |- (DB0.in 3)
      (CP2.out) -| ++(0.60, 0) |- (DB0.in 4)
      (CP1.out) -| ++(0.60, 0) |- (DB0.in 5)
      (CP0.out) -| (DB0.in 6)
      ;

\draw (CP9.out) -| (DB2.in 1)
      (CP8.out) -| ++(0.60, 0) |- (DB2.in 2)
      (CP7.out) -| ++(0.60, 0) |- (DB2.in 3)
      (CP6.out) -| ++(0.75, 0) |- (DB2.in 4)
      (CP5.out) -| ++(0.95, 0) |- (DB2.in 5)
      (CB3.out) -| (DB2.in 6)
      ;

\draw (P3.out) to [short, -*] ++(0.15cm, 0) coordinate(P3_out);

\draw (CT3.in 2) to (P3_out |- CT3.in 2) to (P3_out |- CB3.in 1) to (CB3.in 1);
\draw (AT3.in 2) to [short, -*] (P3_out |- AT3.in 2);
\draw (AB3.in 1) to [short, -*] (P3_out |- AB3.in 1);

\draw (CP1.in 5) to ++(-0.2, 0) to ++(0, -2.75) coordinate(CP1_WIRE) to [short, -o] (base_label |- CP1_WIRE) ++(-0.75, 0) node{$A<B$};
\draw (DB1.in 3) to [short, -*](cp0_cp9 |- DB1.in 3) to [short, -o] (base_label |- DB1.in 3) ++(-0.75, 0) node{$A=B$};
\draw (CP8.in 1) to ++(-0.2, 0) to ++(0, 2.75) coordinate(CP8_WIRE) to [short, -o] (base_label |- CP8_WIRE) ++(-0.75, 0) node{$A>B$};

\draw
      (CT3.in 1) -- ++(-9, 0) |- (AB3.in 2)
      
      (BA3.out) -- ++(0.75, 0) coordinate(BA3_CP5) |- (CP5.in 1)

      (AB0.in 1) -- ++(-0.2, 0) coordinate(AB0_CP3) |- (CP2.in 2)
      (AB0.in 2) |- ++(-3.15, 0) |- (CP2.in 1)

      (P0.out)   to [short, -*](AB0_CP3 |- P0.out)
      (AT0.in 2) to [short, -*](AB0_CP3 |- AT0.in 2)

      (AT2.in 1) -- ++(-4, 0) |- (CP5.in 3)

      (CB3.in 2) |- ++(-9.2, 0) |- (AT3.in 1)

      (CP3.in 1) -- ++(-6.7, 0) |- (AB1.in 2)
      (CP3.in 3) to [short, -*](BA3_CP5 |- CP3.in 3)
      (CP3.in 4) |- ++(-0.4, 0) coordinate(CP3_CP6) |- (CP6.in 1)

      (CP2.in 3) to [short, -*](BA3_CP5 |- CP2.in 3)
      (CP2.in 4) to [short, -*](CP3_CP6 |- CP2.in 4)
      (CP2.in 5) -- ++(-0.75, 0) coordinate(CP2_CP7) |- (CP7.in 1)

      (CP1.in 1) to [short, -*](BA3_CP5  |- CP1.in 1)
      (CP1.in 2) to [short, -*](CP3_CP6  |- CP1.in 2)
      (CP1.in 3) to [short, -*](CP2_CP7  |- CP1.in 3)

      (CP0.in 1) to [short, -*](BA3_CP5  |- CP0.in 1)
      (CP0.in 2) to [short, -*](CP3_CP6  |- CP0.in 2)
      (CP0.in 3) to [short, -*](CP2_CP7  |- CP0.in 3)
      (CP0.in 4) to [short, -*](CP1_WIRE |- CP0.in 4)
            
      (CP4.in 1) |- ++(-9.5, 0) coordinate(CP4_AB2) |- (AB2.in 2)
      (CP4.in 2) |- ++(-6.55, 0) coordinate(CP4_CP5) |- (CP5.in 2)
      (CP4.in 3) to [short, -*](BA3_CP5 |- CP4.in 3)

      (CP6.in 2) to [short, -*](BA3_CP5 |- CP6.in 2)
      (CP6.in 3) -- ++(-7, 0) coordinate(CP6_CP3) |- (CP3.in 2)
      (CP6.in 4) -- ++(-9.5, 0) |- (AT1.in 1)

      (CP7.in 2) to [short, -*](CP3_CP6 |- CP7.in 2)
      (CP7.in 3) to [short, -*](BA3_CP5 |- CP7.in 3)
      (CP7.in 4) to [short, -*](AB0_CP3 |- CP7.in 4)
      (CP7.in 5) -- ++(-9.75, 0) coordinate(P0_CP7) to [short, -*](P0_CP7 |- P0.in 1)

      (CP9.in 2) to [short, -*](CP3_CP6  |- CP9.in 2)
      (CP9.in 3) to [short, -*](CP2_CP7  |- CP9.in 3)
      (CP9.in 4) to [short, -*](CP8_WIRE |- CP9.in 4)
      (CP9.in 5) to [short, -*](BA3_CP5  |- CP9.in 5)
      
      (AT0.in 1) to [short, -*](P0_CP7 |- AT0.in 1)

      (P1.out) to [short, -*](CP6_CP3 |- P1.out)
      (AT1.in 2) to [short, -*](CP6_CP3 |- AT1.in 2)
      (AB1.in 1) to [short, -*](CP6_CP3 |- AB1.in 1)

      (P2.out) to [short, -*](CP4_CP5 |- P2.out)
      (AT2.in 2) to [short, -*](CP4_CP5 |- AT2.in 2)
      (AB2.in 1) to [short, -*](CP4_CP5 |- AB2.in 1)

      (BA2.out) to [short, -*](CP3_CP6 |- BA2.out)
      (BA0.out) -- ++(0.2, 0) coordinate(BA0_CP2) |- (CP1.in 4)
      (BA1.out) |- (BA1.out |- CP8.in 4) -- (BA0_CP2 |- CP8.in 4)

      (CP8.in 2) to [short, -*](CP3_CP6 |- CP8.in 2)
      (CP8.in 3) to [short, -*](CP2_CP7 |- CP8.in 3)
      (CP8.in 4) to [short, -*](BA0_CP2 |- CP8.in 4)
      (CP8.in 5) to [short, -*](BA3_CP5 |- CP8.in 5)

      (DB1.in 1) to [short, -*](CP3_CP6 |- DB1.in 1)
      (DB1.in 2) to [short, -*](BA3_CP5 |- DB1.in 2)
      (DB1.in 4) to [short, -*](CP2_CP7 |- DB1.in 4)
      (DB1.in 5) to [short, -*](BA0_CP2 |- DB1.in 5)
;

\filldraw[fill=yellow!80!black,semitransparent] (-1,3.3) rectangle (1,-11);
\filldraw[fill=green!80!black,semitransparent] (-1,-11) rectangle (1,-18);
\filldraw[fill=red!80!black,semitransparent] (-1,-18) rectangle (1,-24.5);

\filldraw[fill=yellow!80!black,semitransparent] (1.75,3.3) rectangle (3.75,-7);
\filldraw[fill=green!80!black,semitransparent] (1.75,-7) rectangle (3.75,-18);
\filldraw[fill=red!80!black,semitransparent] (1.75,-18) rectangle (3.75,-24.5);

\filldraw[fill=yellow!80!black,semitransparent] (4,3.3) rectangle (6,-10);
\filldraw[fill=green!80!black,semitransparent] (4,-10) rectangle (6,-18);
\filldraw[fill=red!80!black,semitransparent] (4,-18) rectangle (6,-24.5);

\filldraw[fill=yellow!80!black,semitransparent] (8,3.3) rectangle (10,-5.6);
\filldraw[fill=green!80!black,semitransparent] (8,-5.6) rectangle (10,-14.8);
\filldraw[fill=red!80!black,semitransparent] (8,-14.8) rectangle (10,-24.5);

\filldraw[fill=yellow!80!black,semitransparent] (11.5,3.3) rectangle (13.5,-8);
\filldraw[fill=green!80!black,semitransparent] (11.5,-8) rectangle (13.5,-13);
\filldraw[fill=red!80!black,semitransparent] (11.5,-13) rectangle (13.5,-24.5);

\node[draw] at (15.5, -2.35) (a){};
\draw [dashed] (13.5,-2.35) -- (a);
\filldraw[fill=yellow!80!black,semitransparent] (14.5, 1) rectangle (20.25, -5.5);
\draw node[inner sep=0pt, left=-4cm of a] (gpu1){
\includegraphics[width=.25\textwidth,trim={0 0 27cm 0cm},clip]{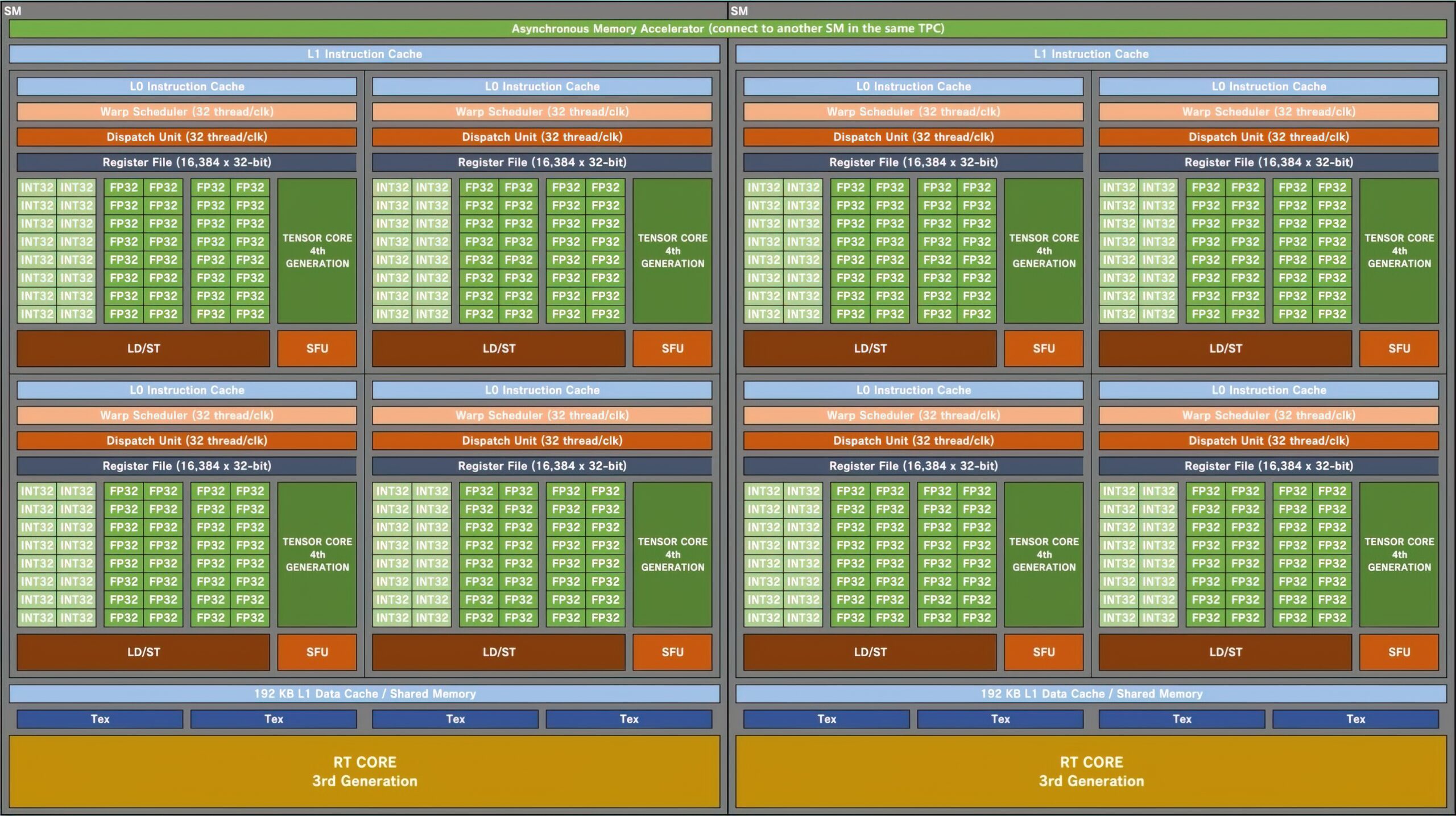}
};

\node[draw] at (15.5, -10.5) (b){};
\draw [dashed] (13.5,-10.5) -- (b);
\filldraw[fill=green!80!black,semitransparent] (14.5, -7.15) rectangle (20.25, -13.65);
\draw node[inner sep=0pt, left=-4cm of b] (gpu2){
\includegraphics[width=.25\textwidth,trim={0 0 27cm 0cm},clip]{figures/gpu_arch.jpg}
};

\node[draw] at (15.5, -18.75) (c){};
\draw [dashed] (13.5,-18.75) -- (c);
\filldraw[fill=red!80!black,semitransparent] (14.5, -15.4) rectangle (20.25, -21.9);
\draw node[inner sep=0pt, left=-4cm of c] (gpu3){
\includegraphics[width=.25\textwidth,trim={0 0 27cm 0cm},clip]{figures/gpu_arch.jpg}
};
\end{circuitikz}
   \caption{\textbf{Mapping gates to devices}: 
    Gates are dispatched in independent waves, which are split across different processing units. In this example, we extract 5 waves which are spread over 3 GPUs that receive a similar workload.}\label{f:dispatch}
\end{figure}

This algorithm terminates even if the circuit is not connected. As the IR can be processed sequentially one node after the other, there cannot be cycles in the circuit and all gates will be given an index.
Since nodes are assigned to waves with indexes that are strictly greater than their predecessors, all entries in a wave are independent and can be processed concurrently, as long as the waves are processed in order. We thus automatically derived a parallel execution from the IR, based on the fact that the IR was a valid sequential execution and used node operands to compute dependencies.
This algorithm has a linear complexity because each node is taken exactly once from the FIFO, and we decrement counters as many times as there are wires in the circuit.
Partitioning the circuit into such waves provides concurrency which can be exploited to efficiently use a single GPU device. For multiple processing units, we dispatch waves over the different devices. A simple solution to dispatch a wave with $N$ gates over $K$ devices that consists of splitting it into roughly $N/K$ gates per device, is illustrated in Figure~\ref{f:dispatch}.

Let us consider a wave with 43875 gates, composed of 2125 \verb+AND+ gates, 25000 \verb+OR+ gates and 16750 \verb+NOT+ gates. On two devices we could have 1356, 10465 and 769 gates of type \texttt{AND}, \texttt{OR} and \texttt{NOT} for device 0; and have 769, 14535 and 6633 gates of these types on device 1. This represents a total of 21938 gates on device 0, and 21937 on device 1, but we measured that device 0 and 1 respectively need 2.21~ms and 2.87~ms to process their portion of the wave. This 30\% load imbalance is explained by the fact that \texttt{AND} and \texttt{OR} gates take the same time to process (about 0.19~us per gate), while it also takes 0.19~us to process 1024 \texttt{NOT} gates, which are non-bootstrapped. Equally dividing the gates between devices is therefore not a satisfactory approach, and only results in a 1.7x speedup with two devices.

We could consider the disparities between the different types of gates to evenly divide the load between the different devices based on performance models, but it would require an extra training phase per gate type. This may be tedious and not reliable when combining multiple gates with different compute or memory bandwidth intensity. In practice, the number of gates is usually large enough that a simpler but effective solution is to assign the same number of identical gates on all processing units. In our previous example, this results in putting 1064 and 1063 \texttt{AND} gates respectively on devices 0 and 1, and putting 12500 \texttt{OR} gates and 8375 \texttt{NOT} gates on both devices 0 and 1. We then measure 2.54~ms of work on both devices, with a negligible difference of less than 0.2~us, which corresponds to a perfect balancing.

For each wave, \algoName implements this strategy using a \emph{hash-table} which associates a vector to each of the opcodes encountered in the wave. Each entry of the wave is then appended to the list which corresponds to its opcode. Considering that there are only 8 types of standard logic gates currently supported, and that this number would not grow significantly, appending an entry roughly has a constant complexity. This phase therefore also has an overall linear complexity. In Section~\ref{sec:backend} we will show that building such vectors of identical gates makes it straightforward to implement batched kernels which obtain much higher performance.

In this Section, we have shown how \algoName converts the frontend IR into a well-balanced parallel workload. Provided CPUs with sufficient processing power, nothing prevents us from assigning them parts of the waves too. This could be done using performance models based on per-gate performance models, or more simply based on the respective peak performance of the different types of processing units. Our methodology is therefore suitable to address hybrid systems combining CPUs and GPUs.

%
%

\section{A fully asynchronous cryptographic backend} \label{sec:backend}
In the previous Section, we described the circuit as a sequence of \emph{waves} subdivided into smaller sets of homogeneous gates to obtain an efficient load balancing over the different processing units. This section details our native CUDA implementation of the CGGI cryptosystem, and explains how we execute this workload as efficiently as possible thanks to a fully asynchronous implementation.
We will now refer to these sets of concurrent homogeneous gates as \emph{batched gates}. 

\subsection{Memory and Communication Considerations}

Since we have covered how gates are batched for distribution to different processing units, we now describe how we can access data across the entire system.

NVIDIA GPUs have a distinct memory hierarchy that differs in key ways from traditional CPUs. 
Inside a streaming multiprocessor (SM), there is a fast on-chip piece of memory partitioned between an L2 cache, and a resource called \emph{shared memory}. These on-chip memories are much faster than global memory as they are part of the SM itself. In fact, shared memory latency is roughly 100x lower than un-cached global memory latency, provided efficient memory access patterns. Shared memory is allocated per thread block, so all threads in the block have access to the same shared memory.
In the case of the A100 GPU, the combined capacity of on-chip memory per SM is 192 kB.
Global memory is the largest memory (40 or 80~GB for the A100) and resides off-chip, making it the slowest aside from accessing memory on the host \cite{nvidia2020a100}.

Bootstrapping keys have a relatively large size, of approximately 100~MB for 110~bits of security. They cannot fit into GPU L2 caches, but are used for the majority of encrypted gate evaluations. Because of this, we replicate them in the global memory of all devices so that each can access the evaluation keys directly. We note that these keys are constant, and can be accessed concurrently within a device. 

Ciphertexts are processed during circuit evaluation, and may be used simultaneously on different devices, or accessed from the host. We thus store them in pinned host memory, which is memory allocated with \texttt{cudaMallocHost()}). Ciphertexts are then cached into shared memory which is much faster and is located close to the GPU SMs.
These kernels indeed have an extremely high arithmetic intensity and the PCI-e bandwidth consumption is limited, and the large amount of concurrency overlaps transfers with computation.
This was verified experimentally by profiling a kernel that processes 1024 gates using the \verb+ncu+ tool. We observed that it only consumed 19.96 MB/s of ``system memory'' bandwidth, which is orders of magnitude lower than the available PCI-e bandwidth. Using pinned host memory to load the input ciphertexts is thus efficient enough, in spite of its simplicity. 

A similar strategy is to use managed memory (also called \emph{unified memory}, and allocated using \texttt{cudaMallocManaged}). Contrary to pinned memory where devices access host memory directly through the PCI-e bus, managed memory is kept coherent across the entire machine by the means of paging mechanisms. When a page fault occurs, the CUDA driver automatically fetches a valid copy of the page where the fault occurred. Subsequent accesses to the same page will occur at the speed of the memory embedded on the device, until the page is evicted from the device.

Both managed memory and pinned host memory incur a significant overhead per allocation, so that we do not allocate all ciphertexts individually, but group these thanks to pooled memory allocators. This pooling mechanism may introduce false sharing issues, but effectively amortizes allocation overhead, which remains noticeable with pinned host memory, but is several orders of magnitude lower than the time required to evaluate the circuit. Memory pages allocated with managed memory and modified concurrently by multiple devices may bounce from one device to another, and have a severe impact on performance.

In practice, we observe similar performance for an encrypted dot product over 8 GPUs with both strategies. With pinned memory, circuit evaluation takes 14.8~s, compared to 15.1~s with managed memory. Allocating 1~GB of pinned host memory however takes 0.4~s, but is negligible with managed memory. Due to the expected page faults when using managed memory on multiple devices, we observe some slightly imperfect parallelism, while it is flawless with pinned memory. \algoName therefore allows user to store ciphertexts either in host pinned or managed memory, for example depending on the amount of system memory which can limit the availability of pinned memory. All experiments presented in the rest of this paper use pinned host memory.

\subsection{Coordinating multiple devices}

A strawman approach to assign tasks to multiple workers involves having a single producer thread and a set of worker/consumer threads.
When using multiple GPUs, each worker thread will consume an assigned batch from the producer and outsource the computation to a dedicated GPU. This approach is quite simple to implement, but requires numerous synchronizations between CPU threads, which negatively impacts scalability by introducing idle periods on the GPUs when CPU threads fail to provide them computation.

Conversely, a more intuitive method involves utilizing a single host thread that will submit work asynchronously to different devices. 
On each device, we create a pool of CUDA streams, so that we can submit multiple concurrent CUDA kernels on this device. The execution of a single wave therefore consists of taking each individual batched kernel from the wave, selecting a CUDA stream on the device on which our scheduling algorithm assigned the batched kernel (\emph{e.g.,} with a round-robin strategy), and submit the kernel in this CUDA stream.
Since waves must be executed in-order, we need to ensure that the execution of a wave does not start until the previous wave has been fully processed. A simple approach would consist of submitting all kernels in a wave on all devices, and then having the host thread wait for the completion of all work on all devices. Waiting for computation to complete from the host however introduces some inefficiency, as devices become idle during the synchronization phase, until the next phase has been submitted. Any potential load imbalance or jitter on a device may also reflect on other devices which could wait longer than expected to get more work.
Instead of blocking devices, we have therefore implemented a non-blocking synchronization fence primitive which ensures that the work in a CUDA stream cannot start running until all work submitted previously in all other streams has been done. These fences are implemented by the means of CUDA events which are asynchronously inserted in the CUDA streams. After inserting an event in each stream, we insert a non blocking CUDA operation which synchronizes one of our CUDA streams with all of these events. We then insert another event in that stream, and make sure all other streams wait for that event. Event insertions and dependency declarations between an event and a stream can be performed asynchronously, ahead of time, and therefore do not require the host thread to block during the execution. These event-based synchronizations are implemented using hardware features, which is much more efficient that having the host thread block the entire device.
This ensures that successive waves can be executed in order, without ever blocking the submission flow of asynchronous operations, until the very end of the circuit evaluation. With this distribution methodology, we observe a speedup of approximately 12\% over the strawman approach for an encrypted dot product benchmark executed on an NVIDIA DGX A100. This may appear to be a moderate improvement, but more than 99\% of the circuit evaluation is spent executing CUDA kernels. We therefore have a close to optimal scheduling strategy over multiple devices, which is essential for the scalability of \algoName according to Amdhal's law. This also indicates that the latency of result retrieval and synchronizations are almost completely hidden.

\subsection{Batched kernels}
Due to the relatively small size of TFHE ciphertexts (compared to other FHE schemes), it is possible to process many FHE gate operations at the same time on GPUs over a large number of ciphertexts. 
Prior works have either launched a separate kernel for every gate evaluation \cite{cufhe} or allow for ``vectorized'' gates (such as performing a bitwise NAND between two 32-bit ciphertext arrays) \cite{nufhe}. 
Conversely, we observe that a better approach for general computation is to leverage a kernel capable of executing arbitrary numbers of gates of any supported type.
The \algoName backend utilizes a single kernel for each batch of gates that launches with $N$ thread-blocks of 512 threads each, where $N$ indicates the number of gates.
This approach is more performant compared to the cuFHE library that initiates host-to-device and device-to-host transfers for each logic gate.
This allows each worker in the runtime environment to launch a single kernel for each batch received from the coordinator, avoiding additional kernel launch overheads. 
Additionally, this technique also allows the GPU to determine the best utilization strategy for the SMs, instead of relying on the user to distribute gates on a per SM basis. 
Grouping gates into homogeneous gates saves memory bandwidth as we only copy the opcode value once per batched kernel, and the generated code is more regular and requires less registers, increasing the \emph{occupancy} of our CUDA kernels~\cite{cuda}. 

Designing batched CUDA kernels which do not require blocking the submitting host thread is also challenging. These kernels indeed need to access buffers with the description of the work to perform, such as the location of the input ciphertexts. We therefore adopt a strategy which consists of assigning such a buffer to each CUDA stream of our pool, and fills them asynchronously from the host using a \emph{host callback}. As a result, our asynchronous batched kernels consist of 1) selecting a CUDA stream on the target device, 2) submitting a host callback that will update the buffer associated to this stream, and 3) launching a CUDA kernel in this stream which will process the batch described using this buffer.
Assigning each CUDA stream a unique buffer requires a limited memory footprint, and ensures there is no concurrent buffer update. This also avoids the use of relatively expensive asynchronous allocations around all asynchronous kernels.

\subsection{NTT Implementation Details}
The performance of CGGI bootstrapping is largely determined by the DFT used to perform polynomial multiplication. 
Both nuFHE \cite{nufhe} and cuFHE \cite{cufhe} use the NTT for this operation, and employ the same strategy in terms of NTT parameters. 
We opt to use these parameters as well, since they provide multiple key optimizations that reduce the NTT latency. 
First, we utilize the modulus $Q = 2^{64}-2^{32}+1$, which simplifies modular reduction and supports NTTs up to size $2^{32}$. 
Lastly, we use the primitive element $g = 12037493425763644479$ that allows most multiplications in the NTT algorithm to become bitshifts modulo $Q$.

\section{Experimental Evaluation} \label{sec:experiments}
\begin{figure*}[!ht]
    \centering
    \begin{subfigure}[b]{0.5\textwidth}
        \centering
        \includegraphics[height=1.3in]{images/dot_product.png}
    \end{subfigure}%
    ~
    \begin{subfigure}[b]{0.5\textwidth}
        \centering
        \includegraphics[height=1.3in]{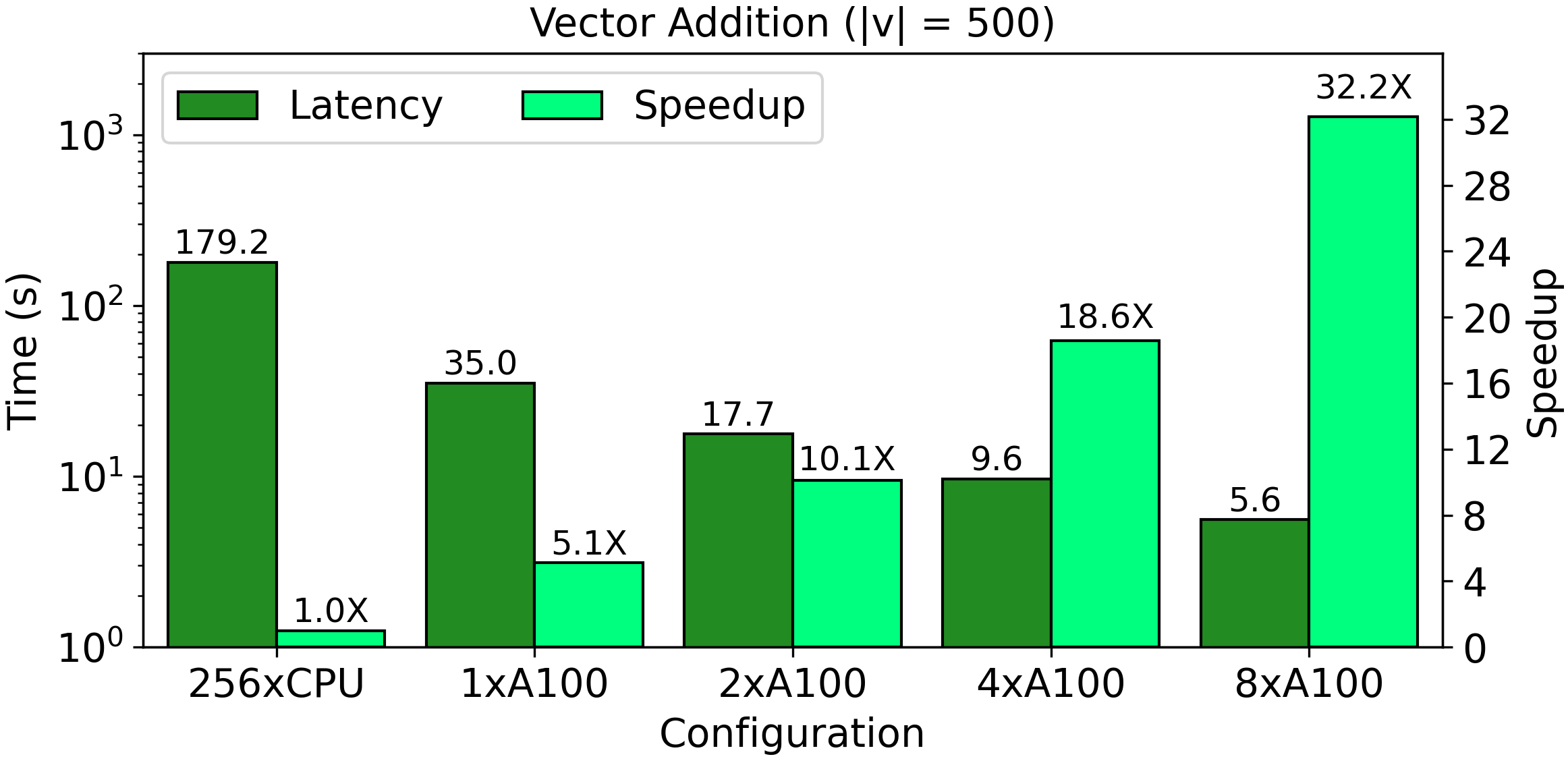}
    \end{subfigure}
    \caption{\textbf{Vector Algebra Benchmarks}: All dot products are performed with 16-bit encrypted elements and the vector addition is performed with 32-bit elements. The speedup bars are relative to the CPU implementation with 256 threads. $|v|$ indicates the vector length and $M$ refers to the dimensions of the matrices.}\label{f:blas}
\end{figure*}

\begin{figure*}[!ht]
    \centering
    \begin{subfigure}[b]{0.5\textwidth}
        \centering
        \includegraphics[height=1.3in]{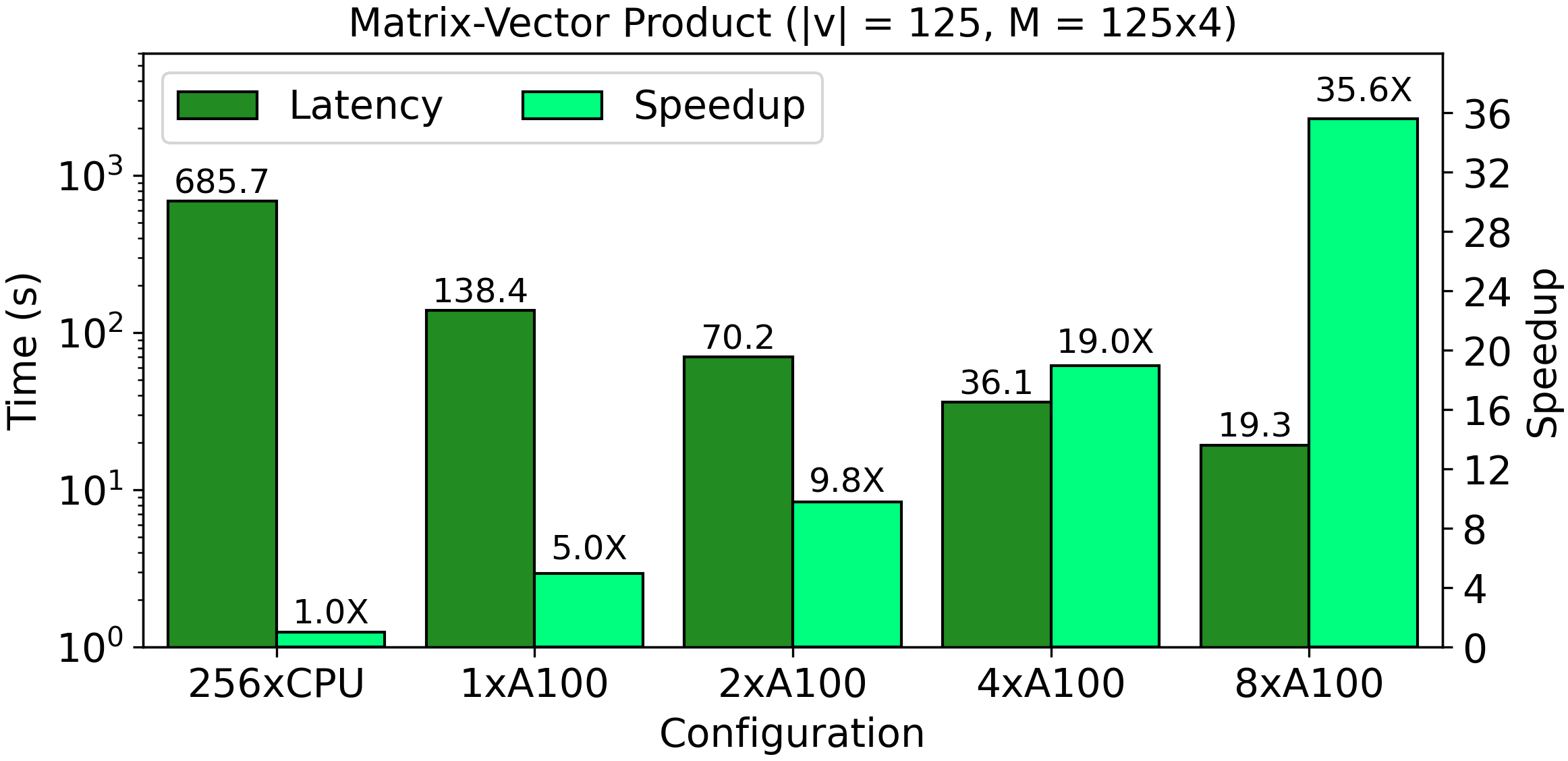}
    \end{subfigure}%
    ~
    \begin{subfigure}[b]{0.5\textwidth}
        \centering
        \includegraphics[height=1.3in]{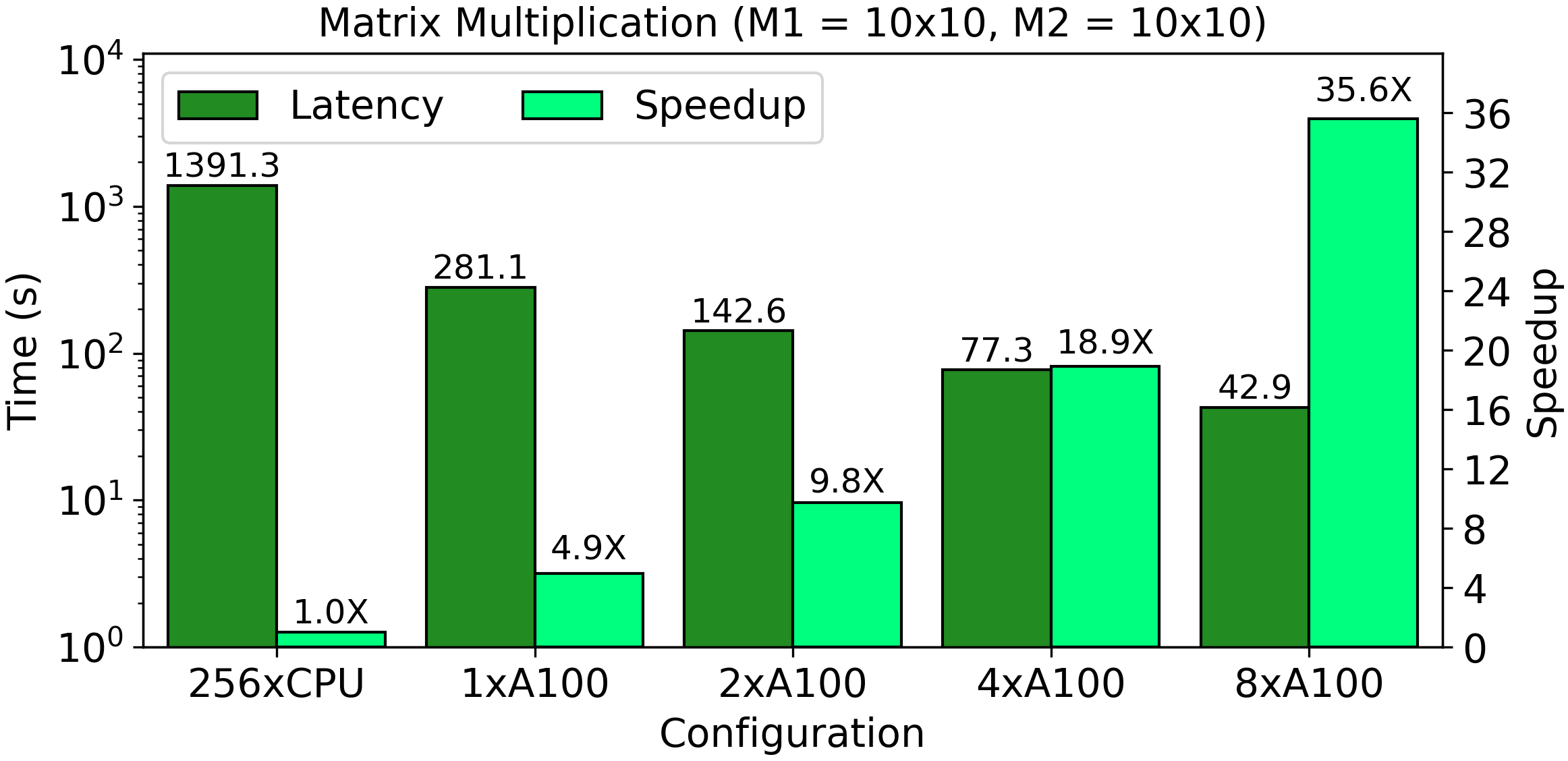}
    \end{subfigure}
    \caption{\textbf{Matrix Algebra Benchmarks}: All products use 16-bit encrypted elements. The speedup bars are relative to the CPU implementation with 256 threads. $|v|$ indicates the vector length and $M$ refers to the dimensions of the matrices.} \label{f:blas_2}
\end{figure*}

\begin{figure*}[!htb]
    \centering
    \begin{subfigure}[b]{0.5\textwidth}
        \centering
        \includegraphics[width=\columnwidth]{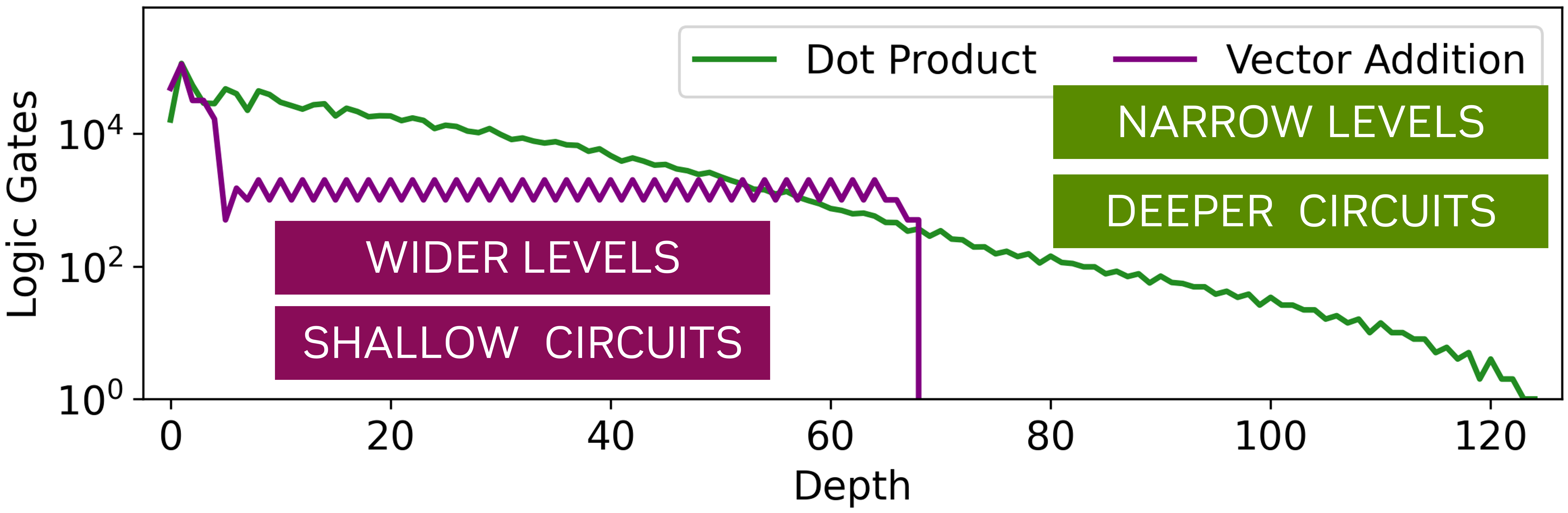}
    \end{subfigure}%
    ~
    \begin{subfigure}[b]{0.5\textwidth}
        \centering
        \includegraphics[width=\columnwidth]{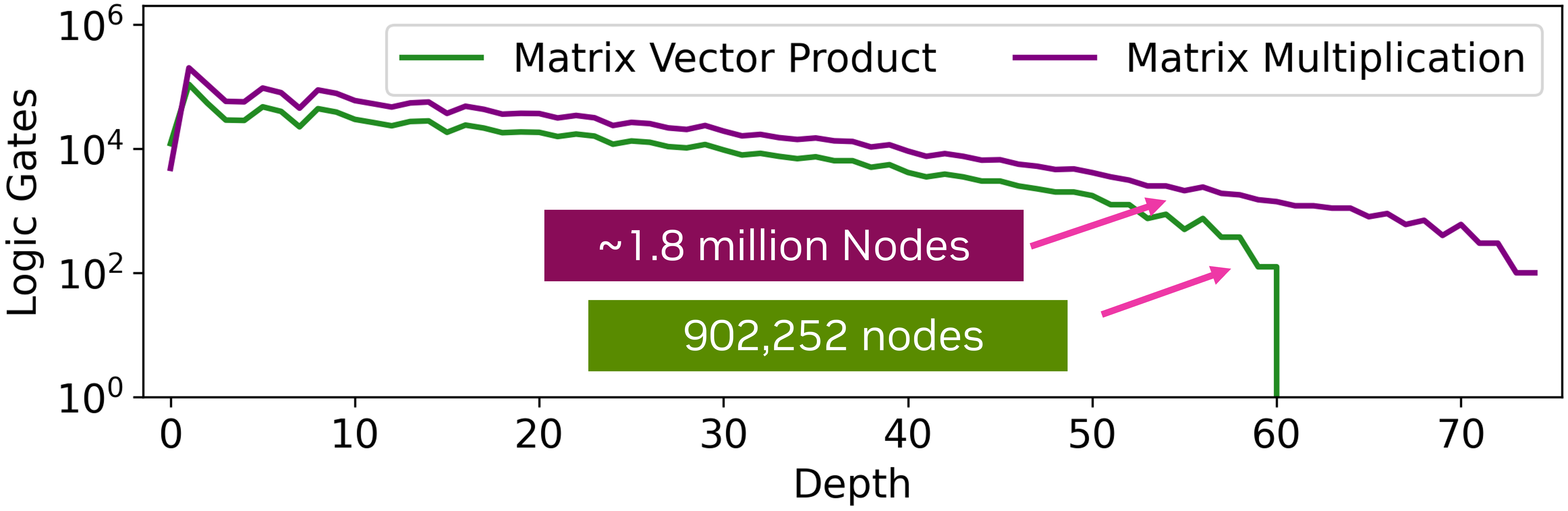}
    \end{subfigure}
    \caption{\textbf{Topology of Linear Algebra Benchmarks}: Vector addition is better suited for circuit for encrypted evaluation as it exhibits wide levels and a short critical path. The matrix-vector product and matrix multiplication benchmarks exhibit ample parallelism, with matrix multiplication being consistently twice as wide for most levels.}\label{f:blas_topology}
\end{figure*}

We employ a series of benchmarks representing realistic computational workloads with FHE to demonstrate the efficacy of \algoName, encompassing areas such as privacy-preserving machine learning, linear algebra applications, and cryptographic benchmarks. 
All experiments were run on an NVIDIA DGX A100, which consists of 8 A100 GPUs and a dual-socket AMD EPYC 7742 CPU with 64 cores each (a total of 128-cores running 256 threads with simultaneous multithreading).
Unless otherwise indicated, all benchmarks were run with parameters corresponding to 110 bits of security based on the BKZ-beta classical cost model provided by the state-of-the-art LWE estimator framework \cite{albrecht2015concrete}. 
Specifically, for RLWE ciphertexts used in bootstrapping, we utilize a ring dimension of 1024 and set the noise rate to $25\times10^{-9}$. 
These are the same RLWE parameters employed by the TFHE library \cite{chillotti2020tfhe} for their parameter set corresponding to 128 bits of security. 
For LWE ciphertexts, we utilize $n = 512$ and a noise rate of $2^{-15}$, which yields approximately 110 bits of security. 
As such, the overall security of the parameter set used for the following experiments is 110 bits of security.
This cost model is also employed by the TFHE library in the security analysis of its hard-coded parameter sets \cite{chillotti2020tfhe}.

\subsection{FHE Basic Linear Algebra Subroutines}
The FHE Basic Linear Algebra Subroutines are benchmarks that form core components of algorithms in a wide variety of fields, such as image processing and  machine learning.
We focus on three distinct tensor multiplication algorithms on 16-bit encrypted data: a dot product of two vectors of length 500, a matrix-vector multiplication between a vector of length 125 and a $125\times4$ matrix, and a matrix multiplication between two $10\times10$ matrices. 
Additionally, we include a vector addition between two vectors of length 500; this benchmark was executed with a larger wordsize than the previous ones to increase its computational complexity.
We compare a $256$-thread CPU execution of these tensor algorithms with our approach running on up to 8 GPUs. 

In Figure \ref{f:blas} and Figure \ref{f:blas_2}, the dark-green bars show running time, and the light green bars plot the speedup of GPU vs. CPU .
One A100 is $5.9\times$ faster than the CPU reference running on the 256-threaded CPU execution model, and $8$ A100s are $40\times$ faster.
We show the latency of these circuits for an increasing number of A100 GPUs and the speedup for all GPU configurations versus a CPU configuration with 256 threads.
Our analysis shows a linear speedup with more GPUs, as our design exploits the ample circuit-level parallelism in both synthesis and runtime phases. 

\begin{figure*}[!ht]
    \centering
    \begin{subfigure}[b]{0.5\textwidth}
        \centering
        \includegraphics[height=1.3in]{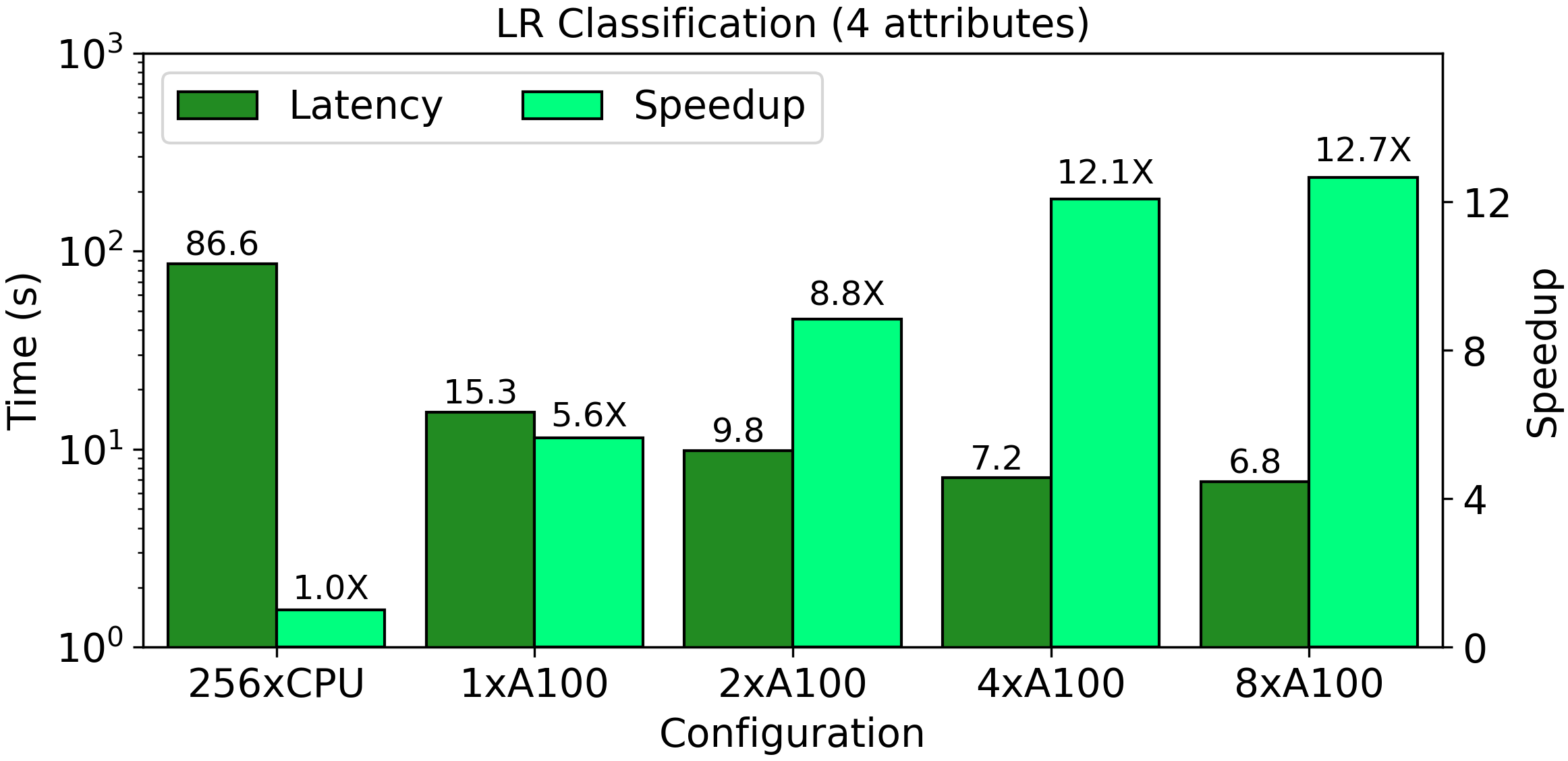}
    \end{subfigure}%
    ~
    \begin{subfigure}[b]{0.5\textwidth}
        \centering
        \includegraphics[height=1.3in]{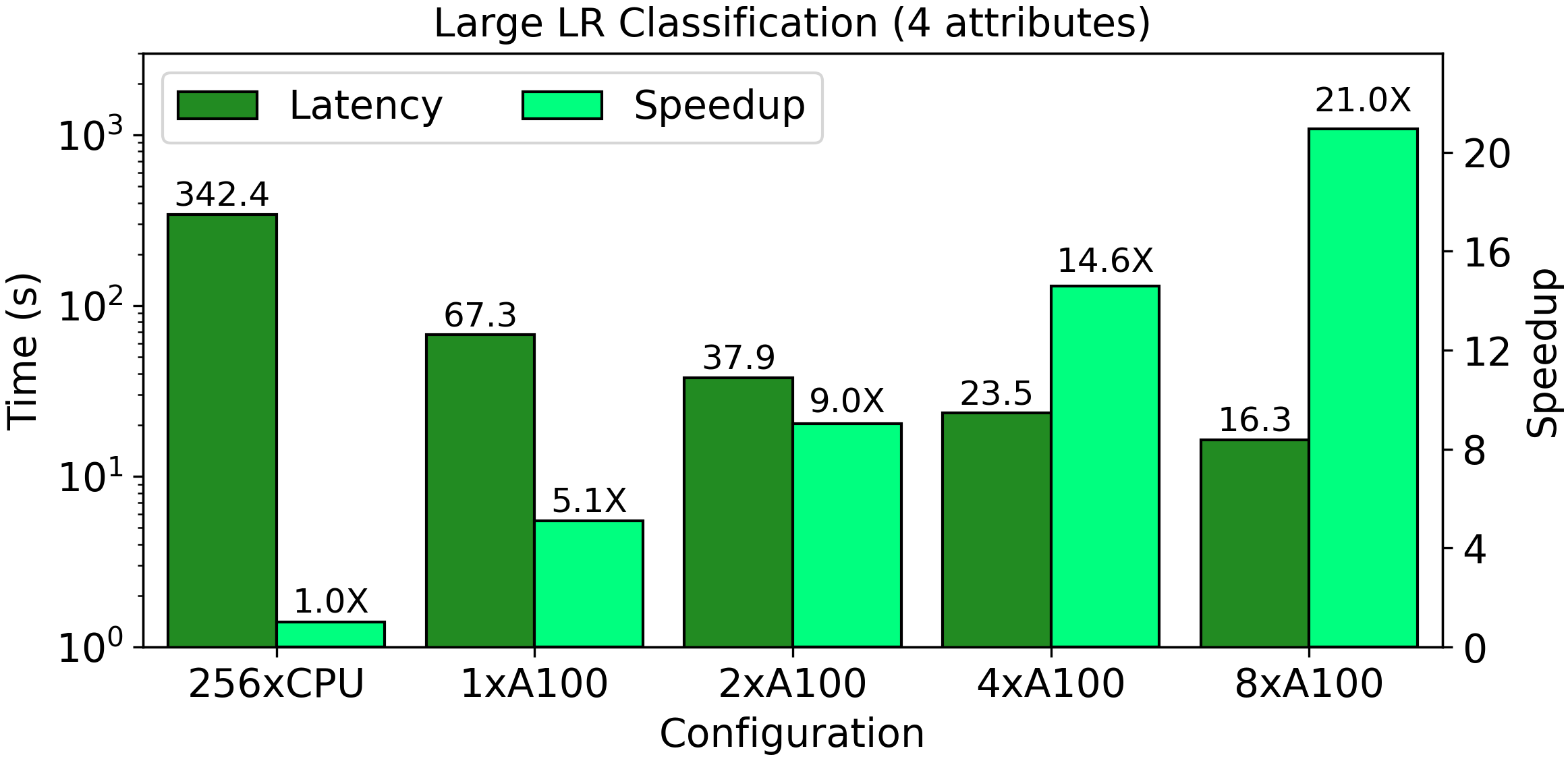}
    \end{subfigure}
    \caption{\textbf{Logistic Regression Inference}: We employ 32-bit words for the small approximation and a 64-bit words for the large approximation to avoid overflow. We observe a better scaling trend for the higher accuracy LR because it exhibits wider levels. }\label{f:lr_inference}
\end{figure*}

\begin{figure*}[!ht]
    \centering
    \begin{subfigure}[b]{0.5\textwidth}
        \centering
        \includegraphics[height=1.3in]{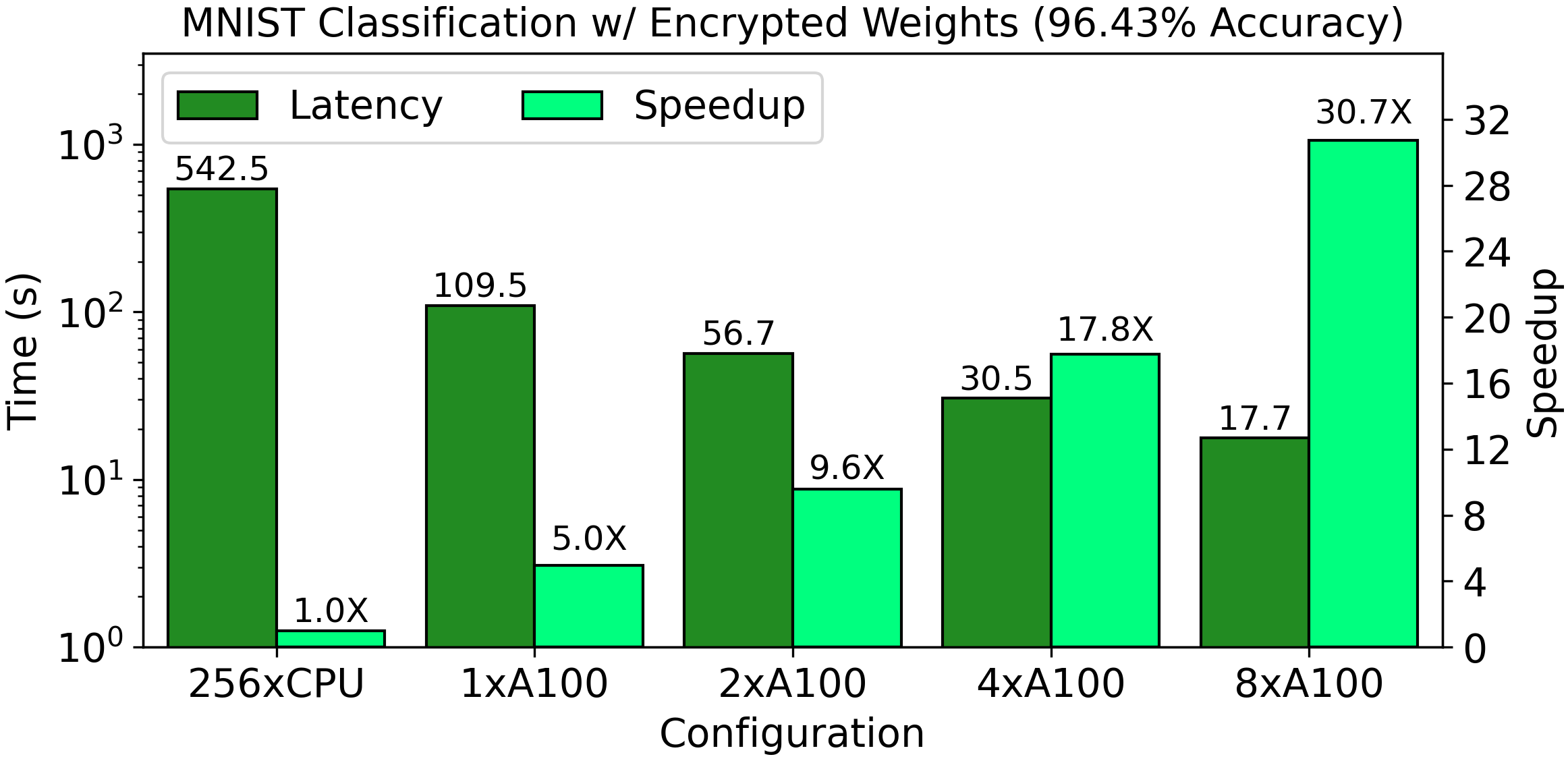}
    \end{subfigure}%
    ~
    \begin{subfigure}[b]{0.5\textwidth}
        \centering
        \includegraphics[height=1.3in]{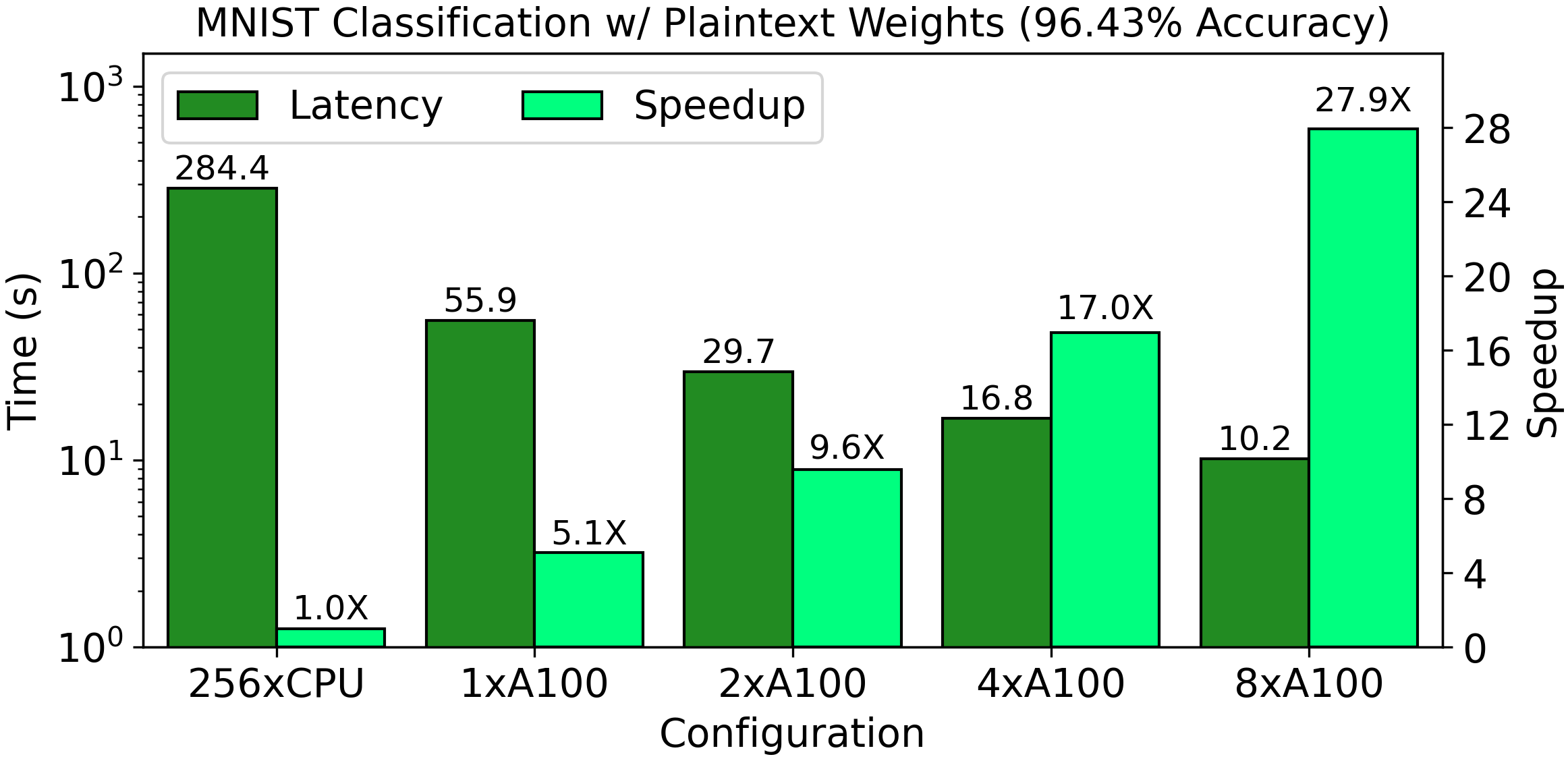}
    \end{subfigure}
    \caption{\textbf{Neural Network Inference}: The encrypted weight variant represents the scenario where the computing party does not own the model, unlike the variant with plaintext weights. We observe a roughly $2\times$ speedup when plaintext weights are used.}     \label{f:nn_inference}
\end{figure*}

\begin{figure*}[!htb]
    \centering
    \begin{subfigure}[b]{0.5\textwidth}
        \centering
        \includegraphics[width=\columnwidth]{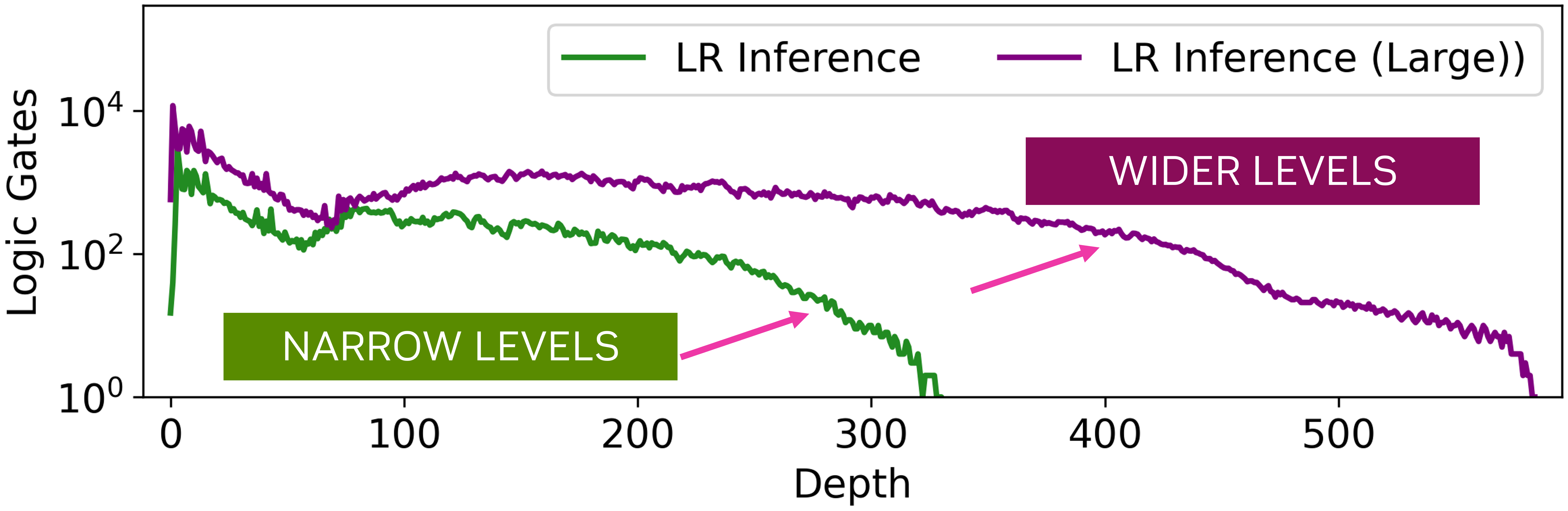}
    \end{subfigure}%
    ~
    \begin{subfigure}[b]{0.5\textwidth}
        \centering
        \includegraphics[width=\columnwidth]{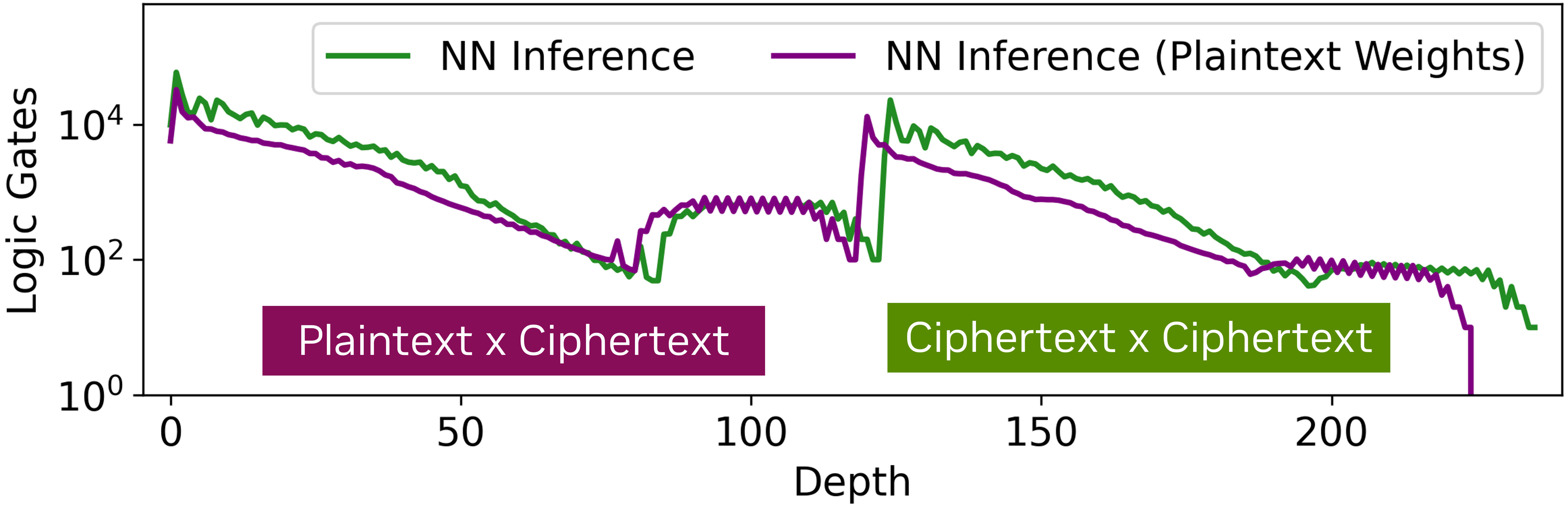}
    \end{subfigure}
    \caption{\textbf{Topology of Machine Learning Benchmarks}: For LR inference, the large variant uses a more accurate sigmoid approximation. It is much deeper due to a larger word size and more polynomial terms evaluated. The neural network plaintext weight variant exhibits a shorter critical path and is composed of much fewer gates overall. }\label{f:ml_topology}
\end{figure*}

Figure \ref{f:blas_topology} shows the linear algebra topologies; the vector addition is more performant as the critical path is approximately $2\times$ shorter and the levels remain wide.
Indeed, this is reflected in the execution times in Figure \ref{f:blas}, where the vector addition runs nearly $4\times$ faster on 8 GPUs. 
Both matrix benchmarks have very wide levels and are well-suited for evaluation on multi-GPU systems. 

\begin{figure*}[!ht]
    \centering
    \begin{subfigure}[b]{0.5\textwidth}
        \centering
        \includegraphics[height=1.3in]{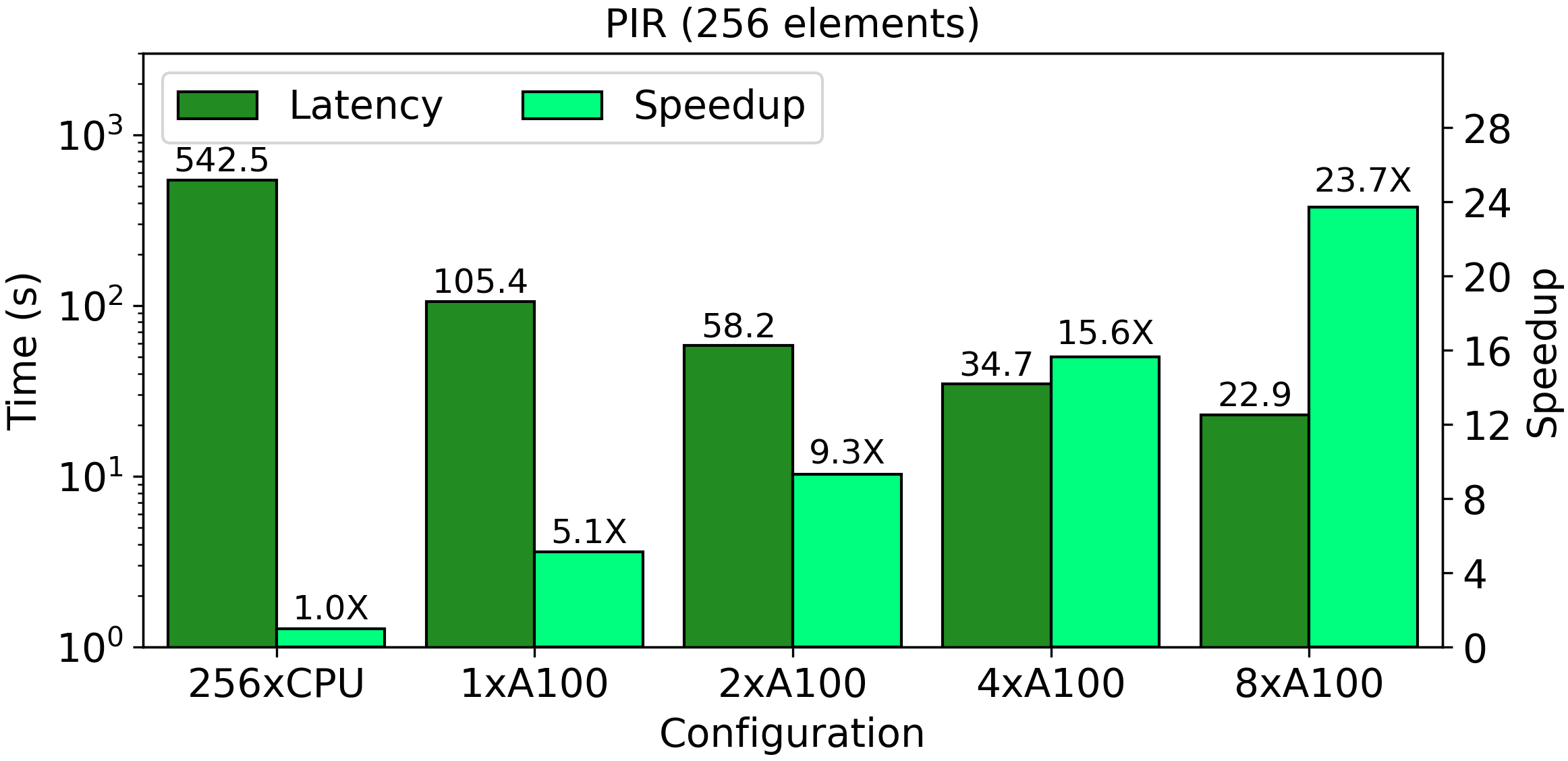}
    \end{subfigure}%
    ~
    \begin{subfigure}[b]{0.5\textwidth}
        \centering
        \includegraphics[height=1.3in]{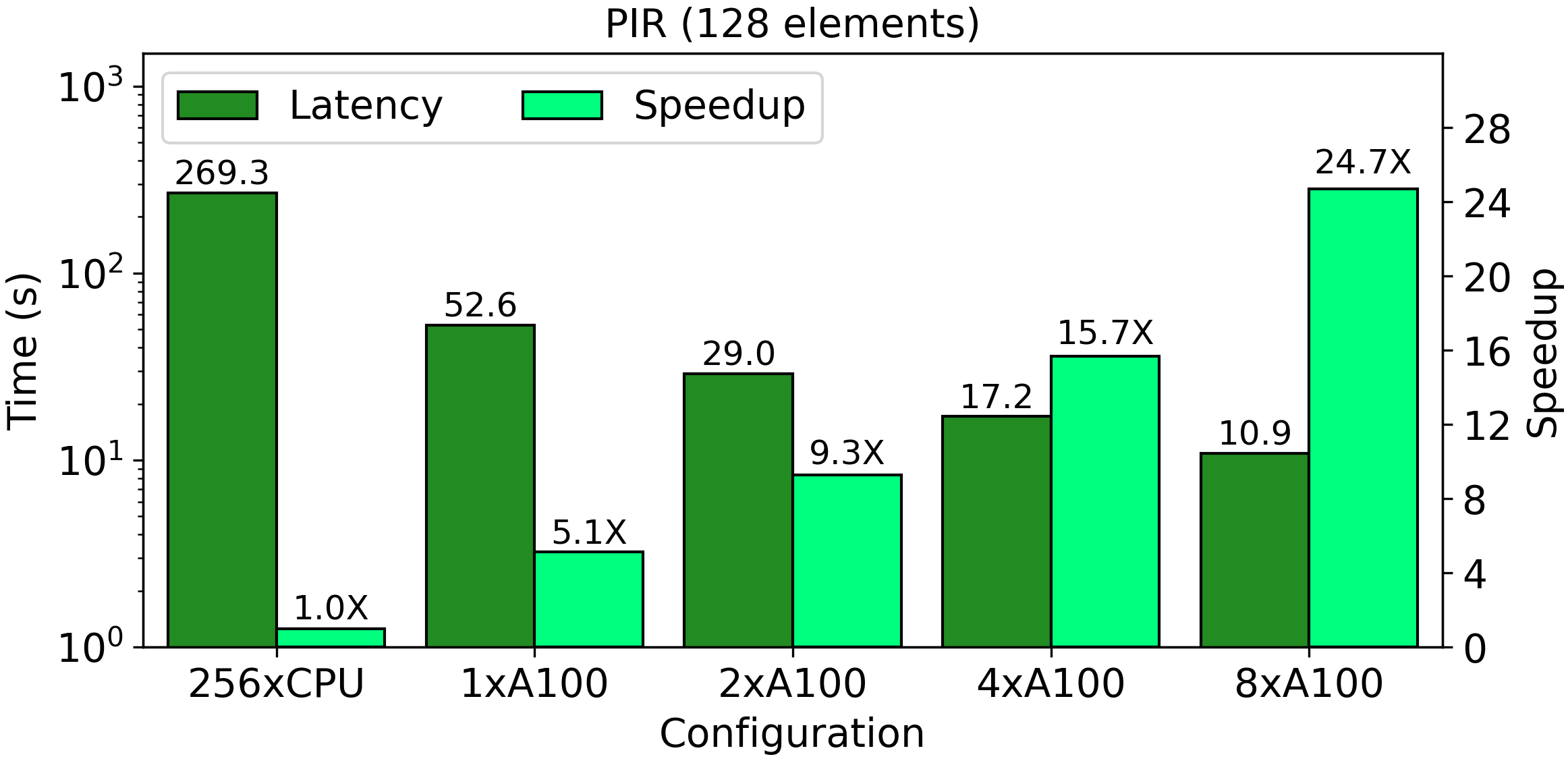}
    \end{subfigure}
    \caption{\textbf{Private Information Retrieval}: We measure the query cost over a key-value storage with 64-bit words.}     \label{f:pir}
\end{figure*}

\begin{figure}[!ht]
    \centering
    \includegraphics[width=\columnwidth]{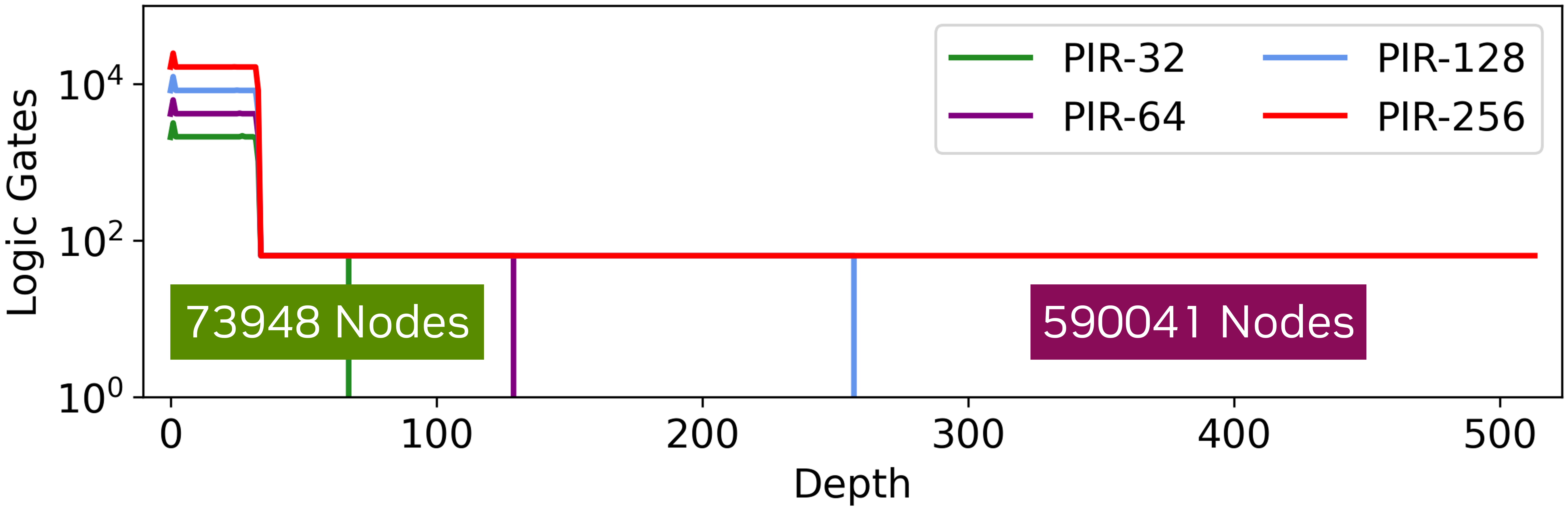}
    \caption{\textbf{Private Information Retrieval Topology}: Increasing the size of the database linearly increases the critical path, and widens the first few layers of the circuit.}\label{f:pir_topology}
\end{figure}

\subsection{Encrypted Machine Learning Applications} 
One of the most widely explored use-cases for FHE is privacy-preserving machine learning as a service.
We consider two distinct scenarios: one where the model is owned by the cloud server and another where the client or a third party owns the ML model. 
For the former, the network parameters are kept in plaintext form and can benefit from faster plaintext-ciphertext operations. 
Conversely, when the cloud doesn't own the model, the network parameters are encrypted as well as the classification inputs.
The majority of existing works demonstrating ML inference with FHE adopt the first scenario as it is generally much more efficient \cite{folkerts2021redsec,bourse2018fast,chou2018faster}. 



Our analysis considers two important machine learning procedures for encrypted classification in the form of logistic regression (LR) inference and neural network (NN) inference. 
For logistic regression, we perform binary classification for datasets with four attributes, such as the Iris dataset \cite{fisher1936use}. 
The key bottleneck in encrypted LR inference is approximating the sigmoid function ($\frac{1}{1+e^{-x}}$), since it is not possible to evaluate it directly. 
Therefore, we employ a polynomial approximation by evaluating terms of the Maclaurin series.
In general, when approximating nonlinear activation functions, there is a trade-off between accuracy and computational complexity.
We show this trade-off with \algoName through the use of an approximation that evaluates the first four terms, and one that evaluates the first six.
Figure \ref{f:lr_inference} shows diminishing returns when increasing the number of GPUs due to the large critical path and relatively thin circuit levels of the benchmark in Figure \ref{f:ml_topology}.
Using 8 GPUs with the more accurate sigmoid approximation still outperforms the CPU implementation with 256 threads by $21\times$. 

For neural network inference, we employ the same network architecture used by FHE-DiNN \cite{bourse2018fast}.
The network consists of two fully-connected layers with a sign activation function to classify the MNIST dataset of handwritten digits. 
We consider two variants of this network that differ in whether or not the network parameters are encrypted and both achieve an accuracy of 96\% for MNIST classification.
The execution times across both configurations are depicted in Figure \ref{f:nn_inference} while Figure \ref{f:ml_topology} presents the characteristics of these workloads. As expected, the variant where the cloud server does not own the proprietary network (i.e., using encrypted weights) has approximately $2\times$ higher latency because of the increased number of ciphertext-ciphertext operations. 

\subsection{Private Information Retrieval}

Aside from machine learning and linear algebra benchmarks, we explore another realistic and useful application of FHE enabled by \algoName in the form of private information retrieval (PIR).
The ability to search and perform computation across an encrypted database has many useful applications, such as managing a directory of health-care records that must be kept confidential for compliance with standards such as HIPAA \cite{zhang2017searchable}. 
We represent the encrypted database as a key/value storage where both keys and values are encrypted.
Figure \ref{f:pir_topology} demonstrates the circuit characteristics of a single query for databases of increasing size.
For brevity, we show results for databases with 256 and 128 elements, where each value is 64-bits in size.
In Figure \ref{f:pir}, we observe a linear scaling with increasing database size; this is expected as a query requires comparison operations with each record in the database due to the \textit{termination problem}. 
Specifically, the termination problem states that it is impossible to make a decision based on encrypted data as the computing party does not know the underlying value of the ciphertext \cite{mouris2018terminator}.
As such, each element of the database must be visited and the correct entry needs to be chosen through oblivious encrypted multiplexing operations.

\section{Related Works} \label{sec:related}
Prior works can be divided into two categories: FHE compilers for general-purpose computation and GPU frameworks that reduce the latency or improve throughput of homomorphic operations. 
The former category targets the usability issue inherent in FHE and explores automatic application-level optimizations to facilitate efficient execution for the target backend. 
The \algoName frontend and middle-layer address these challenges as well, and can be directly compared prior works in this line of research. 
The latter category includes works that focus on FHE acceleration using both software and hardware techniques at the primitive level, and are also comparable to our proposed backend. 

\subsection{FHE Compilers}
The Cingulata framework (formerly Armadillo \cite{carpov2015armadillo}) allows users to map C++ code into a sequence of \texttt{AND} and \texttt{XOR} gates.
Cingulata works strictly with binary FHE contexts using the TFHE library (which implements CGGI) and a custom BFV implementation as its backends. 
Compared to \algoName, Cingulata only supports single-core execution for CGGI and does not offer GPU support. 
The BFV mode is parallelized on CPUs, but does not support bootstrapping and hence cannot be used for arbitrary general-purpose computation. 

E$^{3}$ is a C++ library that introduces custom encrypted data types for leveraging FHE in general applications \cite{chielle2018e3}. 
It supports a variety of cryptographic backends, including TFHE, Microsoft SEAL, and HElib, encompassing all major FHE schemes. 
Unlike \algoName, E$^{3}$ uses a direct mapping to hardcoded FHE functional units and does not offer an optimizing compiler. It also does not support any GPU-friendly cryptographic backends and no parallelization is included.

Google's FHE Transpiler \cite{gorantala2021general} and Romeo \cite{gouert2020romeo} leverage logic synthesis and optimizations to generate FHE programs for general computation. 
However, both works employ generic synthesis scripts that include optimizations not relevant to encrypted computation.
The FHE Transpiler targets TFHE and the OpenFHE implementations of the CGGI cryptosystem as backends, and can evaluate multiple gates in parallel using interpreter mode. 
However, it does not support GPUs and its parallelization strategy is not suited for them, yielding very low device utilization. 
Likewise, Romeo  targets TFHE and generates an FHE program instead of interpreting it. 
This approach, however, does not scale for large programs or HPC systems as described in Section \ref{sec:middleware}. Conversely, \algoName offers a novel dispatch strategy and multigate kernels that can efficiently compute batches of any set of gates.

\begin{table}[!ht]
\centering
\caption{Comparisons with existing backends for 32-bit arithmetic operations (taken from Morshed et al. \cite{morshed2020cpu}). Other backends are configured for 80 bits of security, while ArctyrEX is configured for 110 bits of security. All backends run on a single GPU besides the CPU-based TFHE that runs on a single thread. For the GPU frameworks, a technology scaling factor was introduced for fair comparisons (defined as the number of SMs in our A100 GPU divided by the number of SMs of the target GPU).}
\begin{tabular}{@{}ccc@{}}
\toprule
\textbf{Library} & \textbf{Add (s)} & \textbf{Mult (s)} \\ \midrule
\algoName & \textbf{1.33} & \textbf{2.13} \\
FHE Transpiler~\cite{gorantala2021general} & 6.53 & 13.56 \\
Morshed et al.~\cite{morshed2020cpu} & 1.47 & 25.13 \\
TFHE~\cite{chillotti2020tfhe} & 7.04 & 489.93 \\
nuFHE~\cite{nufhe} & 3.08 & 137.78 \\
cuFHE~\cite{cufhe} & 1.50 & 97.5  \\ 
REDcuFHE~\cite{folkerts2021redsec} & 1.55 & 99.21 \\ \bottomrule
\end{tabular}\label{t:comparisons}
\end{table}

\subsection{FHE Acceleration Frameworks}
The cuFHE \cite{cufhe} and nuFHE \cite{nufhe} constitute the current state-of-the-art for GPU acceleration of the CGGI cryptosystem. 
The former is a proof-of-concept library that implements high throughput logic gate evaluations on a single NVIDIA GPU. 
However, cuFHE is not configurable (i.e., only supports 80 bits of security), has non-optimal data transfers and requires frequent high-cost synchronization between GPU and CPU. 
Each cuFHE gate evaluation requires all ciphertext inputs be copied from the host to the device, and each output is copied back from the device to the host. 
This approach is impractical for realistic circuit evaluation, as it yields millions of large ciphertext transfers between the CPU and GPU. 
Lastly, not all cuFHE computations are outsourced to the GPU and the CPU needs to perform certain operations (such as evaluating the homomorphic \texttt{NOT} gate). 
Unfortunately, this defeats the benefits gained from asynchronous CUDA kernel launches and the CPU execution must block when it reaches a \texttt{NOT} gate until the GPU has finished evaluating all prior gates, instead of continuing to do more meaningful work. 
Similarly, nuFHE specializes in vectorized gates; for instance, it can evaluate a bitwise \texttt{AND} operation across 64-bit operands. 
However, this approach is very restrictive in terms of circuit evaluations as typically a circuit level is not composed of one type of gate. 

REDcuFHE \cite{folkerts2021redsec} enhances cuFHE to add multi-bit plaintext support and multi-GPU support. 
However, it still suffers from the same synchronization issues as cuFHE, and puts the burden of scheduling and handling communication between multiple GPUs on the programmer. 
\algoName, on the other hand, handles all scheduling and communication procedures automatically. 
Lastly, Morshed et al. \cite{morshed2020cpu} present a GPU implementation of CGGI that leverages the NVIDIA cuFFT library and incorporates a set of handwritten circuits such as vector addition and matrix multiplication. 
Table \ref{t:comparisons} demonstrates that \algoName outperforms \cite{morshed2020cpu} by a factor of about $1.5\times$ for a small 32-bit addition and $16\times$ for 32-bit multiplication (which is a significantly larger circuit).
Additionally, \algoName evaluates a vector addition with 32 elements of 32-bit integers $4.1\times$ faster and a $16\times16$ matrix multiplication of 32-bit elements $10.6\times$ faster.
We also emphasize that all frameworks in Table \ref{t:comparisons}, aside from \algoName and the Google FHE Transpiler, require developers to write their own circuits by hand, as opposed to automatically generating them. 

A final class includes custom ASICs and FGPA implementations of FHE to act as a co-processor to speed up the underlying FHE arithmetic operations. However, many of these designs only support limited parameter sets \cite{nabeel2022cofhee, turan2020heaws, roy2019fpga, samardzic2021f1} and usually target other cryptosystems that enable approximate computing \cite{samardzic2022craterlake, agrawal2023fab} or computation with modular integers \cite{nabeel2022cofhee}. On the other hand, ArctyrEX can support arbitrary parameters and implements the CGGI cryptosystem, which is more suitable for general-purpose computation.

\section{Concluding Remarks} \label{sec:conclusion}
\algoName is an end-to-end framework for general-purpose encrypted computation that leverages GPU acceleration and incorporates novel strategies for executing FHE algorithms efficiently.
For workloads such as neural network inference, we observe a linear speedup with increasing GPUs thanks to the inherent circuit-level parallelism, the proposed dispatch paradigm, and the high degree of primitive-level parallelism exploited by our CUDA-accelerated CGGI backend. 

In future work, we plan to expand our frontend to support schemes beyond CGGI, as different schemes are better suited to different styles of computation.
For instance, computing multiplications with large word sizes in CGGI is inefficient because the underlying circuit will be very large. Other schemes, like CKKS and BGV, support encrypting multi-bit values directly and can accomplish this multiplication in one primitive operation.
Moreover, CKKS is an attractive option for certain machine learning applications, as it natively supports operations on encrypted floating-point values. With small modifications to our current \algoName frontend, namely omitting the logic synthesis step, we can readily support all other FHE schemes that take the form of a general arithmetic circuit as opposed to Boolean circuits.

Developments in our middleware layer involves investigating further scheduling optimizations that further reduce device-to-device data transfers. A potential solution to this challenge involves incorporating graph partitioning methodologies to minimize inter-level dependencies between computing devices.
Regarding our backend, future work will investigate alternative techniques to accelerate the DFT step, such as exploring further NTT acceleration on GPUs, as well as adopting the FFT. 
We also plan to investigate fusing gate evaluations across GPU streaming multiprocessors to minimize latency of FHE gates. 
This capability will be useful for thin circuit levels where the total number of gates is less than that of the total number of SMs across all available GPUs.
For the application level FHE optimization, future research involves integrating deep neural network optimizations such as \cite{DBLP:journals/micro/JosephGMGG20, DBLP:phd/us/Joseph21} and its correctness emphasis \cite{9296941,joseph2020going, dam2023understanding} with optimizations in the\algoName frontend to achieve higher throughput and reliably accurate encrypted deep learning inference.

\newpage

\appendix

\section{Transciphering Benchmarks}

At first glance, it seems odd to compute an encryption algorithm homomorphically when the data is already encrypted. 
However, these algorithms enable an exciting strategy called \emph{transciphering} that dramatically reduces the large communication overhead associated with FHE \cite{albrecht2015ciphers,cho2021transciphering}. 
Instead of sending large homomorphic ciphertexts to the cloud for outsourced computation, the client can send encryptions generated with a traditional block or stream ciphers that result in little to no data expansion. 
Then, the cloud can \textit{homomorphically} decrypt the received symmetric ciphertext by evaluating the corresponding decryption algorithm of the chosen cipher using an homomorphic encryption of the symmetric key. 
For the CGGI cryptosystem at 110 bits of security, this strategy can decrease the communication overhead associated with the client sending encrypted inputs by a factor of over $16000\times$. 

\begin{table}[!ht]
\centering
\caption{Amortized decryption latency for Speck and Simon}
\begin{tabular}[]{ccc}
\hline
\textbf{Configuration} & \textbf{Speck Round (s)} & \textbf{Simon Round (s)} \\ \hline
256xCPU        & 2.41                            & 0.80                                \\
1xA100                 & 0.34                             & 0.13                               \\
2xA100                 & 0.29                             & 0.07                                \\ \hline
\end{tabular}\label{t:crypto}
\end{table}

Our analysis employs the lightweight Simon and Speck ciphers proposed by the US National Security Agency \cite{beaulieu2015simon}. 
These ciphers are well-suited to evaluate CGGI cryptosystems because they are primarily composed of bitwise operations.
Other ciphers like AES are less suitable as they require expensive lookup-table evaluations, or a high number of finite-field arithmetic operations \cite{gentry2012homomorphic}.
For both ciphers we use their 128/128 bit variants, as symmetric security needs to be commensurate to our FHE parameters. 
Table \ref{t:crypto} presents the cost per round to evaluate Simon and Speck per 128-bit block size.
Overall, Simon is more efficient than Speck because it uses strictly bitwise operations, whereas Speck has a 64-bit subtraction in each round that corresponds to a large Boolean circuit. 

\section{Developing FHE Applications}

\algoName is designed to run efficiently for any algorithm, yet the way that an algorithm is expressed can have an impact on the HLS procedures that map the program to a Boolean netlist. 
The most important consideration lies in the complexity of the high-level application code. 
For complex algorithms that contain several thousand loop iterations or large loop bodies, the time required to perform the HLS and logic synthesis can increase significantly, or the synthesis toolchain itself may be unable to successfully generate a circuit. 
Notably, we can overcome this challenge by splitting an algorithm into multiple HLS inputs and invoking them one after the other in the encrypted application. Figure~\ref{f:partialmm} demonstrates this strategy; the kernel implements the inner loop of a $10\times10$ matrix multiplication and can be invoked multiple times to evaluate the full GEMM procedure.

\newsavebox{\partialmm}
\begin{lrbox}{\partialmm}
\begin{lstlisting}[language=C,
    breaklines=true,         
    basicstyle=\ttfamily\footnotesize,
    numbers=left,
    numberstyle=\scriptsize\color{gray},
    keywordstyle=\color{blue},
    commentstyle=\itshape\color{gray}
]
int partial_mm(int x[10], int y[10]) {
  int res = 0;
  for (int i = 0; i < 10; i++) {
    res = res + x[i] * y[i];
  }
  return res;
}
\end{lstlisting}
\end{lrbox}

\begin{figure}[h!]
\centering
\usebox{\partialmm}
\vspace{0.1in}
\caption{Partial Matrix Multiplication HLS Kernel}\label{f:partialmm}
\end{figure}

\section{NVIDIA DGX A100 System}

The NVIDIA A100 DGX used in this work consists of 8 A100 GPUs. 
Two distinct groups of four GPUs are inter-connected using high speed NVLink buses. 
%
These GPUs have hardware support for direct access to \emph{registered} host memory, which we leverage for intermediate encrypted wire transfers before they are cached into the shared memory of the devices during gate evaluation.
Along with the synchronization mechanisms employed by our custom dispatching system, this naturally ensures data consistency across multiple devices.
The NVIDIA A100 GPUs used in our experimental evaluation are data center GPUs, which is consistent with prior works (e.g., \cite{chillotti2020concrete}).
Each GPU has 108 streaming multiprocessors that act as independent processing units.
Each streaming multiprocessor can evaluate a logic gate in the context of \algoName, allowing for 108 concurrent gate evaluations on a single GPU.

\section{Security Considerations and Threat Model}

\algoName generates code for a third-party cloud server to perform computations on encrypted data.
We assume an honest-but-curious computing party, where the server can be trusted to do the expected computation but has incentives to view the sensitive user inputs. 
The server is aware of the underlying size and type of the data being manipulated (for example, integer, string, or class), as well as the evaluated algorithm.
If the length of the data needs to be protected for a given application, we assume this is enforced on the client-side by introducing fixed input lengths.

Our existing backend is based on the CGGI scheme \cite{chillotti2020tfhe} which bases its security on the (R)LWE problems.
In cryptography, the security of a cipher is established using cryptanalysis and the security is derived from a reliance on underlying mathematical problems that are known to be NP-hard.
this is directly applicable to CGGI, as LWE and its variants are all hard lattice problems. 
%





\begin{acks}
Authors acknowledge the members of 
Zama, 
CryptoLab, 
Google and 
Duality Tech 
particularly  
Ilaria Chillotti, 
Jung Hee Cheon, 
Ahmad Al Badawi, 
Yuri Polyakov, 
David Cousins, 
Shruti Gorantala and 
Eric Astor
for their insightful discussions.
We also acknowledge the constructive feedback from the anonymous reviewers and editors, which significantly contributed to the quality of this paper.
\end{acks}

\newpage

\bibliographystyle{ACM-Reference-Format}
\bibliography{main}

\end{document}